\documentclass[journal,transmag]{elsarticle}

\usepackage{setspace}
\usepackage{amssymb,amsmath,amsthm}
\usepackage{verbatim} 
\usepackage{listings} 
\usepackage{color}
\usepackage[normalem]{ulem}

\newcommand{\bb}{\mathbf}

\usepackage{tikz}
\usetikzlibrary{arrows}
\usepackage{pgfplots}

\newproof{pf}{Proof}

\usepackage{algorithm}
\usepackage{algorithmic}

\usepackage{multirow}

\usepackage{enumitem}

\usepackage[margin=0.75in]{geometry}

\begin{document}
	
\begin{frontmatter}

	\title{Spectral Resolution Clustering for Brain Parcellation
	}


\author[newhavenaddress]{Keith Dillon\corref{mycorrespondingauthor}}
\cortext[mycorrespondingauthor]{Corresponding author. Tel. 1-949-478-1736}
\ead{kdillon@newhaven.edu}
\author[tulaneaddress]{Yu-Ping Wang}

\address[newhavenaddress]{Department of Electrical and Computer Engineering and Computer Science, \\ University of New Haven, West Haven, CT, USA}
\address[tulaneaddress]{Department of Biomedical Engineering, \\ Tulane University, New Orleans, LA, USA}


\begin{abstract}
	
We take an image science perspective on the problem of determining brain network connectivity given functional activity. 
But adapting the concept of image resolution to this problem, we provide a new perspective on network partitioning for individual brain parcellation. 
The typical goal here is to determine densely-interconnected subnetworks within a larger network by choosing the best edges to cut. 
We instead define these subnetworks as resolution cells, where highly-correlated activity within the cells makes edge weights difficult to determine from the data.
Subdividing the resolution estimates into disjoint resolution cells via clustering yields a new variation, and new perspective, on spectral clustering. 
This provides insight and strategies for open questions such as the selection of  model order and the optimal choice of preprocessing steps for functional imaging data.
The approach is demonstrated using functional imaging data, where we find the proposed approach produces parcellations which are more predictive across multiple scans versus conventional methods, as well as versus alternative forms of spectral clustering.

\end{abstract}

\end{frontmatter}



\section{Introduction}

The brain is widely understood to operate via  densely-interconnected network behavior \cite{fornito_fundamentals_2016}.
Many diseases have been proposed to be essentially ``connectivity diseases", such as schizophrenia, Alzheimer's, and other dementias \cite{van_den_heuvel_exploring_2010}. 
Hence, initiatives such as the Human Connectome Project \cite{sporns_human_2005} have high hopes for solving open problems in brain function and disease \cite{craddock_imaging_2013}. 
At the same time, modularity of function, to at least some degree, is also clearly evident in the brain.
Brain lesions in specific locations commonly lead to specific defects \cite{blumenfeld_neuroanatomy_2010}, for example Broca's aphasia resulting from lesions in Broca's area, or memory defects from lesions in the hippocampus.  
Another source of support comes from the regional differences in cytoarchitecture, identified by Brodmann and others in cadavers \cite{zilles_centenary_2010}.
Due to this combination of modular and network function, the brain is often described as having a hierarchical architecture \cite{van_den_heuvel_exploring_2010}. 
Brain parcellation, therefore, can be described as identification of the macroscopic-to-mesoscopic level (or levels) of this hierarchy.


The simplest and most common \cite{stanley_defining_2013}, approach to parcellation is the use of pre-defined neuroanatomical maps defining regions of interest (ROI) \cite{evans_brain_2012}.
Functional imaging data, such as a functional magnetic resonance imaging (fMRI) scan, is first normalized into a common coordinate system system such as Talairach coordinates \cite{lancaster_automated_2000}. Then, pre-defined masks provide a direct identification of groups of image voxels onto regions describing the parcels.
From here researchers can compute the net activity of the parcel by averaging the time courses of contained parcels, then analyze the networked activity between the parcels.
However the brain is known to vary significantly between individuals, particularly in association regions of the cortex \cite{wang_parcellating_2015}, rendering the subsequent analyses inaccurate for such regions.
This is of course problematic as it is the complex processes which are believed to involve such regions \cite{blumenfeld_neuroanatomy_2010}, which are the most poorly-understood functions of the brain. 
Further, it is not clear that cytoarchitecture alone can serve to determine the functional divisions in the brain \cite{arslan_joint_2015}.

A variety of data-driven methods for parcellation have been investigated \cite{thirion_which_2014,craddock_neuroimage_2018,de_reus_parcellation-based_2013, arslan_human_2018}.
These use the similarities between the time courses of the voxels to group them into parcels \cite{blumensath_spatially_2013}. 
The simplest idea is to apply clustering methods, unsupervised learning approaches from the machine learning field \cite{hastie_unsupervised_2009}, directly to the voxel time courses.
Multiple types of clustering have been adapted to parcellation, including $k$-means clustering \cite{goutte_clustering_1999, thirion_which_2014}, hierarchical clustering \cite{goutte_clustering_1999,filzmoser_hierarchical_1999,arslan_multi-level_2015,blumensath_spatially_2013}, and fuzzy clustering approaches \cite{baumgartner_fuzzy_1997,baumgartner_quantification_1998,golay_new_1998}. 
Related methods include the identification of boundaries between groups of similar time courses \cite{gordon_generation_2016}, matrix factorization methods \cite{blumensath_sparse_2014,cai_estimation_2018}, dictionary learning \cite{wang_cerebellar_2016}, and processing the time courses to form new types of features to cluster \cite{goutte_feature-space_2001,mezer_cluster_2009}.
Fuzzy c-means was once considered the favorite \cite{goutte_feature-space_2001}, but research has largely taken a different direction in recent decades.
Despite several years of progress, the problem of individual parcellation continues to be considered unsolved, with no clearly-superior approach \cite{arslan_human_2018}.

Recent research has sought to utilize the network connectivity itself to improve upon parcellation \cite{eickhoff_connectivity-based_2015}.  
The most direct approach is to apply a clustering method to the covariance matrix or a similar quantity describing correlations between time courses.
Such analyses have primarily been limited to specific sub-regions such as the orbitofrontax cortex \cite{kahnt_connectivity-based_2012}, post-central gyrus \cite{roca_inter-subject_2010}, or medial frontal cortex \cite{kim_defining_2010}.
This is likely due at least in part to the fact that clustering of correlations produces a poor parcellation at the scale of the entire brain. 
Approaches such as Yeo et al \cite{thomas_yeo_organization_2011} yield small numbers (e.g., seven) of brain-wide networks, rather than spatially-localized parcels. 

%

A closely-related direction is based on the graph partitioning perspective, leveraging advances from spectral graph theory \cite{chung_spectral_1997} and the fast-growing field of network science \cite{brandes_what_2013}. 
Here the time courses are used to generate a large graph connecting all voxels, which may be weighted or binary, though to our knowledge it is always undirected and unsigned.
Then a graph partitioning technique such as normalized cuts (Ncut) \cite{shi_normalized_2000}  or spectral clustering \cite{weiss_segmentation_1999,meila_learning_2002,belkin_laplacian_2003,saerens_principal_2004,von_luxburg_tutorial_2007,meila_spectral_2015}  is applied to this dense graph, defining parcels as subgraphs with denser internal connections.
As with the clustering of correlations, immediate adaptations of these methods  tend to lack the spatial continuity of ROI, instead resulting in a small number of brain-wide networks \cite{heuvel_normalized_2008,venkataraman_exploring_2009}, or being restricted in analysis to a single ROI \cite{shen_graph-theory_2010}.
Some researchers have developed methods which achieve more realistic-looking parcels by imposing spatial constraints \cite{thirion_dealing_2006}, such as by only allowing adjacent pixels to be connected \cite{craddock_whole_2012,arslan_joint_2015}, or by combining neighbors into ``super-voxels" \cite{wang_supervoxel-based_2016,wang_generation_2018}.  
%
%
%
While spatial constraints or similar regularization techniques exhibit good reproducibility, this may simply be a result of bias, as such methods do not necessarily perform well in other metrics \cite{arslan_human_2018}.

In this paper we present a new approach to parcellation based on taking an image science perspective on the problem of predicting activity within a network, as initially proposed in \cite{dillon_regularized_2017}.
In imaging, a resolution cell is defined as the smallest region within which no further detail may be discerned.
Here we analogously consider groups of voxels who's connectivity cannot be independently resolved, and suggest this implies they belong in the same parcel. 
We will show that using clustering to produce a parcellation of resolution cells in this way, yields a new variant on spectral clustering, which is able to form realistic-looking ROI without need for strict distance regularization. 
We demonstrate the approach using real fMRI data, where we find that the resulting parcels are also more predictive of network activity. 

%
%

%
%
%
%
%
%

\section{Methods}
%
%
%
%
%
%
%
%

Spectral clustering is a very popular clustering technique \cite{von_luxburg_tutorial_2007} which has been used to solve problems in a wide range of areas such as image processing \cite{weiss_segmentation_1999}, graph theory \cite{saerens_principal_2004}, clustering on nonlinear manifolds \cite{belkin_laplacian_2003}, and brain parcellation as reviewed in the previous section.
Spectral clustering is commonly described as a continuous relaxation of the normalized cut algorithm \cite{von_luxburg_tutorial_2007} for partitioning graphs.
The discrete version of the normalized cut algorithm is NP-hard \cite{von_luxburg_tutorial_2007}, while the continuous approximation (which we call ``spectral clustering", often also referred to in the literature as simply ``the Ncut algorithm") can be performed with basic linear algebra methods.
The most common form \cite{von_luxburg_tutorial_2007} of spectral clustering is provided in Algorithm 1. 
\begin{algorithm}
\begin{algorithmic}[1]
	\caption{Spectral Clustering}
	\STATE Form undirected unsigned weighted graph represented by weighted adjacency matrix $\bb W$ from correlations between time-courses. Choose number of clusters $K$. 
	\STATE Compute graph Laplacian $\bb L = \bb D - \bb W$. The matrix $\bb D$ is a diagonal matrix where $D_{i,i}$ is the degree of node $i$. 
	\STATE Compute $K$ smallest eigenvalues $\lambda_1, ..., \lambda_K$ and corresponding eigenvectors $\bb (\bb v_1, ..., \bb v_K) = \bb V_1$ from the eigenvalue decomposition $\bb L = \bb U \bb S \bb V^T$, where $\bb v_i$ is the $i$th column of $\bb V$.
	\STATE Apply $k$-means clustering to the rows of $\bb V_1$ with $K$ clusters. 
	\label{SC_algo}
\end{algorithmic}
\end{algorithm}

In addition to approximately-optimal partitioning of graphs, other interpretations have been noted for spectral clustering \cite{meila_spectral_2015}, sometimes involving minor variants on Algorithm \ref{SC_algo}, such as by normalizing the Laplacian differently. 
In \cite{meila_learning_2002}, it is shown that one can also use the largest eigenvalues and corresponding eigenvectors of a normalized version of the adjacency matrix $\bb W$ (normalized to have unit row sums).
The question of which variant works best has generally been left to empirical testing \cite{von_luxburg_tutorial_2007}. 

Our approach addresses the problem from an entirely different direction, based on the idea of resolving unknown edges, leading to a new variant of spectral clustering. 
In \cite{dillon_image_2016} the concept of resolution was applied to estimating the relation of functional imaging to phenotypes, resulting in modular regions suggestive of brain parcellation.
In \cite{dillon_regularized_2017}, the resolution concept was applied to the relation between the activity of different voxels, i.e., network estimation, and used to perform brain parcellation.
First we will review that approach here.

Let $\bb A$ be a matrix containing fMRI data, where $\bb a_i$, the $i$th column of $\bb A$, contains the time series describing the activity of the $i$th voxel. 
We assume the data has been preprocessed to remove artifacts, and standardized.
The neighborhood selection problem \cite{meinshausen_high-dimensional_2006} is the regression problem to estimate the functional connectivity of the $k$th voxel given all other voxels, which we formulate as
\begin{align}
\begin{array}{c}
\bb x_k^* = \arg\underset{\bb x_k}{\min} \; \Vert \bb A \bb x_k  - \bb a_k \Vert,
\end{array}
\label{eq_nbhd_opt}
\end{align}
where $\bb x_k^*$ is the $k$th column of the weighted adjacency matrix $\bb X$.
Note that we do not restrict the diagonal of $\bb X$ to be zero, hence we allow self loops in the network.
A typical approach is to seek a sparse solution to Eq. (\ref{eq_nbhd_opt}), as in \cite{meinshausen_high-dimensional_2006}, by imposing an appropriate regularization term.
This is especially valued in applications such as connectomics due to the high redundancy in the data, i.e., strong correlations between voxel time series. 

\begin{figure}[h!] \centering 
	\scalebox{0.55}{\includegraphics[clip=true, trim=0in 0in 0in 0in]{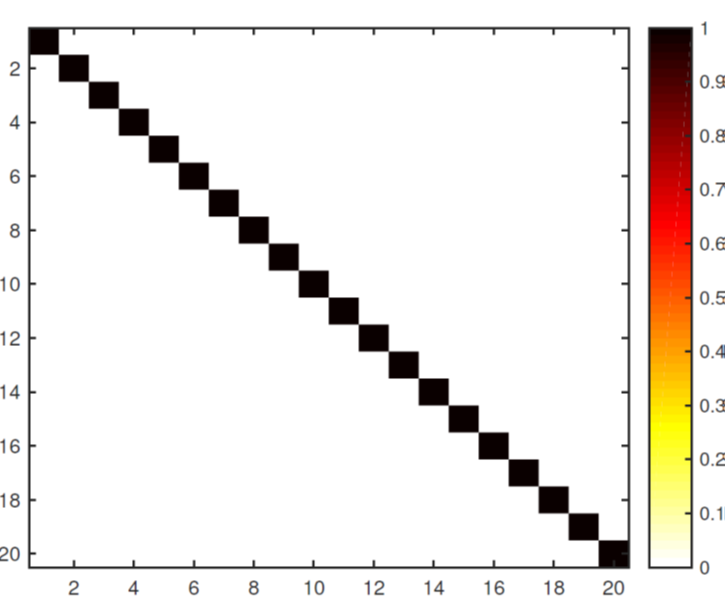}} 
	\scalebox{0.55}{\includegraphics[clip=true, trim=0in 0in 0in 0in]{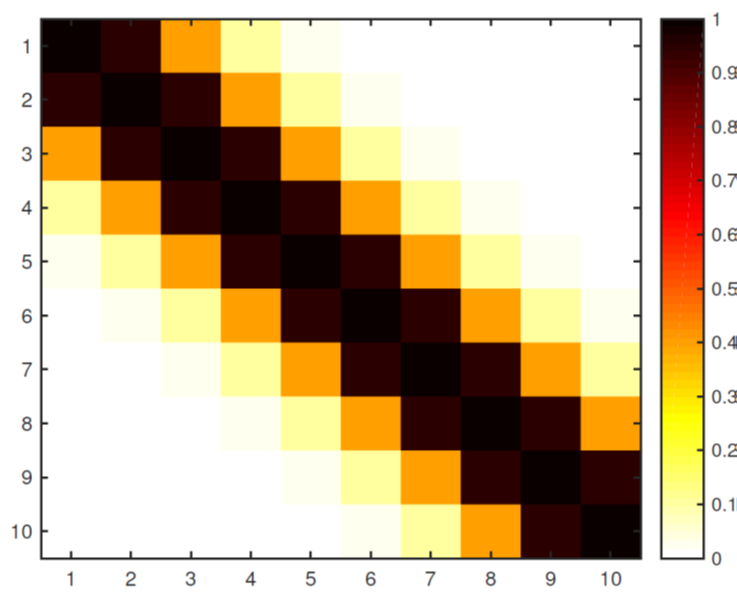}} 
	\caption{Identity matrix (left) versus resolution matrix (right) which depicts a blurring of each sample among the nearest three samples. if the resolution equals the identity then we have maximum resolution, as we can perfectly reconstruct the unknown $\bb x_k$.}
	\label{img_identity_vs_reso}
\end{figure}

Of course the goal of parcellation is to identify modules based on these correlations. 
To that end we will use an approach from image science for analyzing the redundancy in an inverse problem.
In this case, the inverse problem is the linear model,
\begin{align}
	\bb A \bb x_k  = \bb a_k,
	\label{eq_inverse_nbhd}
\end{align}
where we view $\bb a_k$ in Eq. (\ref{eq_nbhd_opt}) as a measured output, and $\bb x_k$ as an unknown input we wish to determine.
The data matrix $\bb A$ serves as a forward model, an operator which transforms the input to the output, causing some degree of information loss. 
The resolution matrix \cite{jackson_interpretation_1972} describes this information loss for inverse problems, and is defined as
\begin{align}
    \bb R = \bb A^\dagger \bb A, 
\end{align}
Where $\bb A^\dagger$ is the pseudoinverse of $\bb A$.  
The resolution matrix is sometimes described as an approximate identity matrix for the problem \cite{ganse_uncertainty_2013}. 
Consider that if $\bb A$ was invertible, then  $\bb A^\dagger = \bb A^{-1}$ and hence $\bb R = \bb I$, the identity matrix. Further, an invertible $\bb A$ means Eq. (\ref{eq_inverse_nbhd}) can be solved exactly for $\bb x_k$ using the inverse, and indeed Eq. (\ref{eq_nbhd_opt}) will find this exact solution. 
Generally, the more $\bb R$ differs from the identity matrix, the more information loss there is, and the worse our estimate of the true connectivity will be.
A simple example is depicted in Fig. \ref{img_identity_vs_reso}.
We can see that this is a kind of minimal choice of blurring region by solving for the  best approximation to an inverse, using the following optimization program
\begin{align}
\begin{array}{c}
\bb y_i^* = \arg\underset{\bb y_i}{\min} \; \Vert \bb A^T \bb y_i  - \bb e_i \Vert_2^2,
\end{array}
\label{eq_resoopt}
\end{align}
where $\bb e_i$ is a vector of zeros with a value of one in the $i$th element. 
The solution to this problem is
\begin{align}
\bb y_i^* = (\bb A^T)^\dagger \bb e_i,
\label{eq_y_pinv}
\end{align}
which is a row of the pseudoinverse.
Further, note that $\bb A^T \bb y_i^* = \bb A^T (\bb A^T)^\dagger \bb e_i = \bb r_i$, the $i$th row of $\bb R$.
So $\bb r_i$ is an optimal approximation to $\bb e_i$.
Since $\bb e_i$ is the $i$th row of the identity matrix, $\bb R$ is an optimal approximation to the identity.

\begin{figure}[h!] \centering 
	\scalebox{0.27}{\includegraphics[clip=true, trim=0in 0.0in 0in 0.0in]{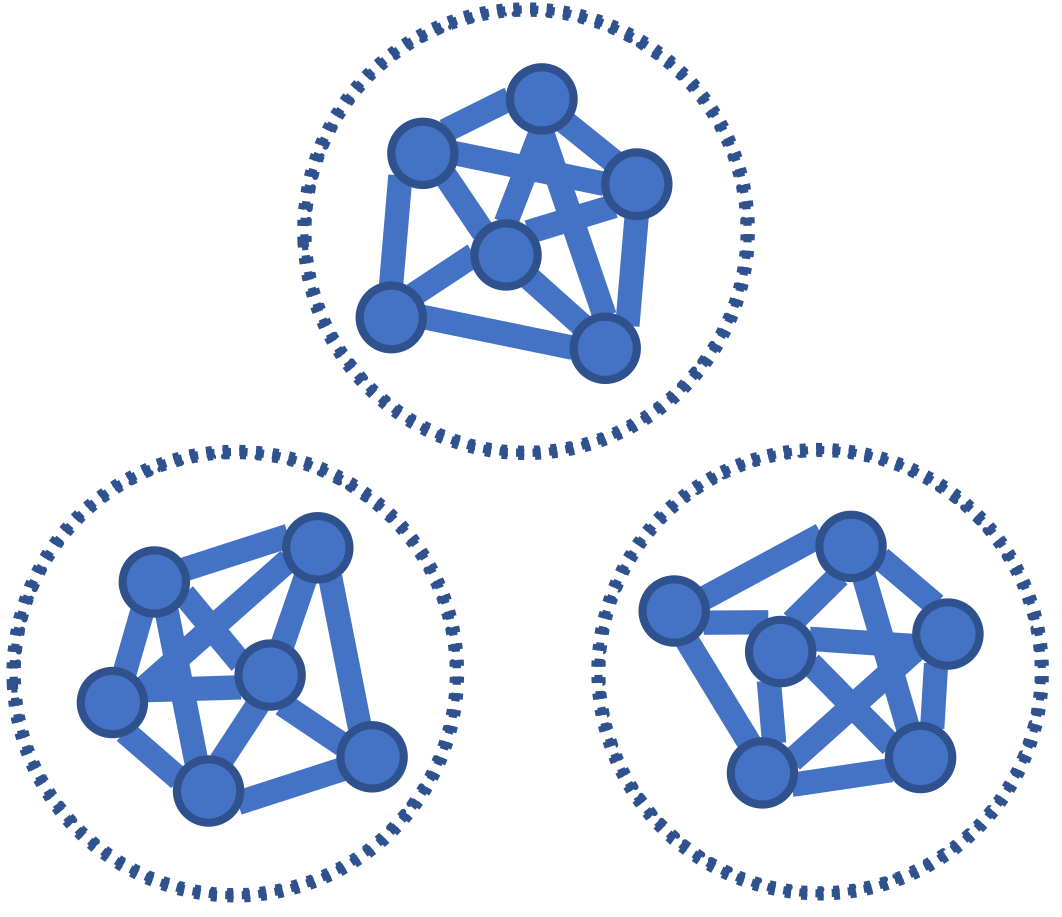}} 
	\scalebox{0.35}{\includegraphics[clip=true, trim=0in 0in 0in 0.15in]{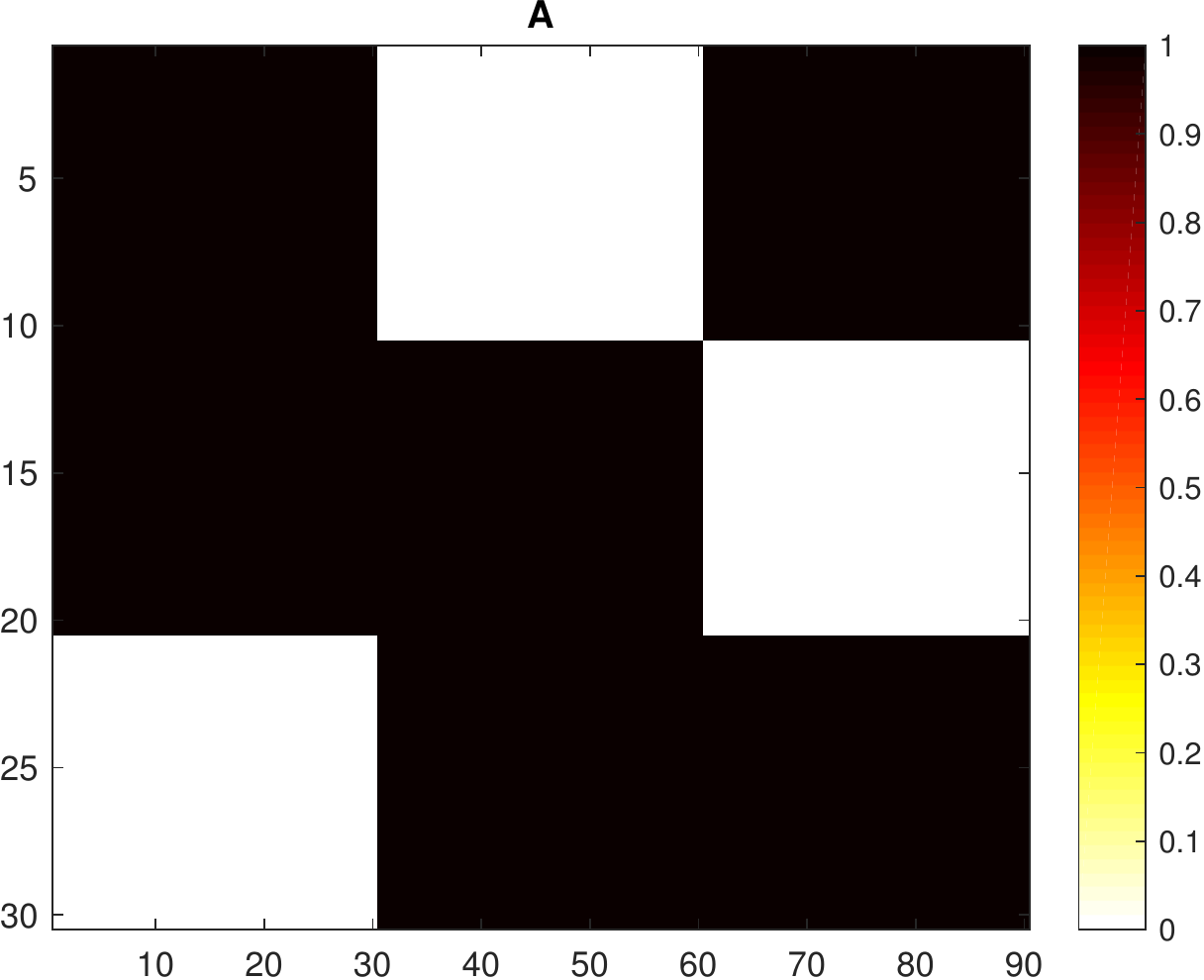}} 
	\scalebox{0.35}{\includegraphics[clip=true, trim=0in 0in 0in 0.15in]{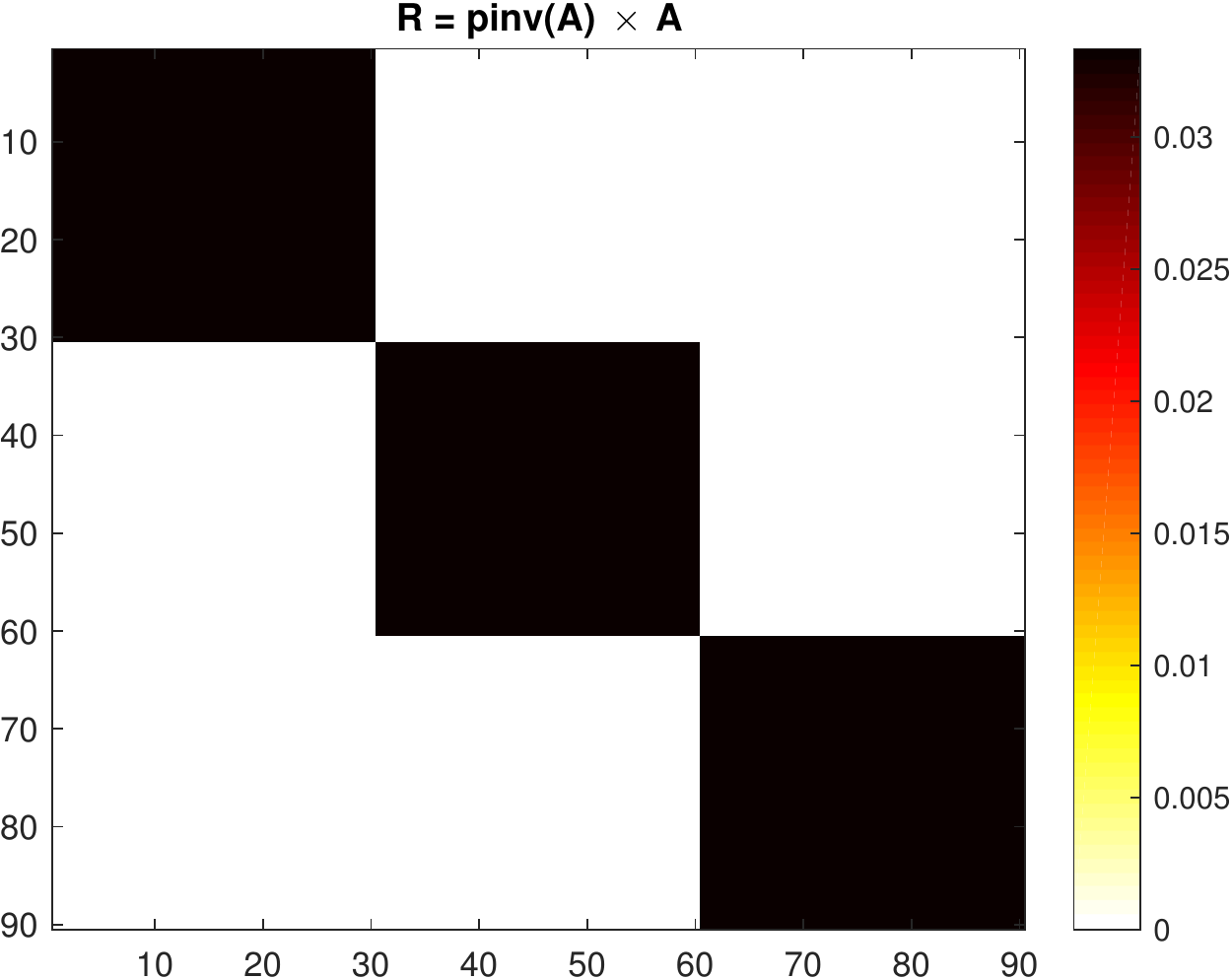}} 
	\caption{Simple network (left) consisting of densely-interconnected nodes in three groups; corresponding data matrix (middle) and resolution matrix (right) which is able to resolve the subnetworks individually, but not the nodes within, as the groups differ in their time courses but the nodes within each group have identical time courses.}
	\label{pic_network}
\end{figure}

In Fig. \ref{pic_network} we give a simple simulation to demonstrate the resolution matrix for a network, where correlation is due to connectivity rather than simply missing information. 
Fig. \ref{pic_network} depicts the network, data matrix of time courses, and resolution matrix. 
We see that the resolution matrix, formed by applying the pseudoinverse of the data matrix to the data matrix itself, is able to separate the three groups but not resolve nodes any further.

\subsection{Regularization}

\begin{figure}[h!] \centering 
	\scalebox{0.35}{\includegraphics[clip=true, trim=0in 0in 0in 0.15in]{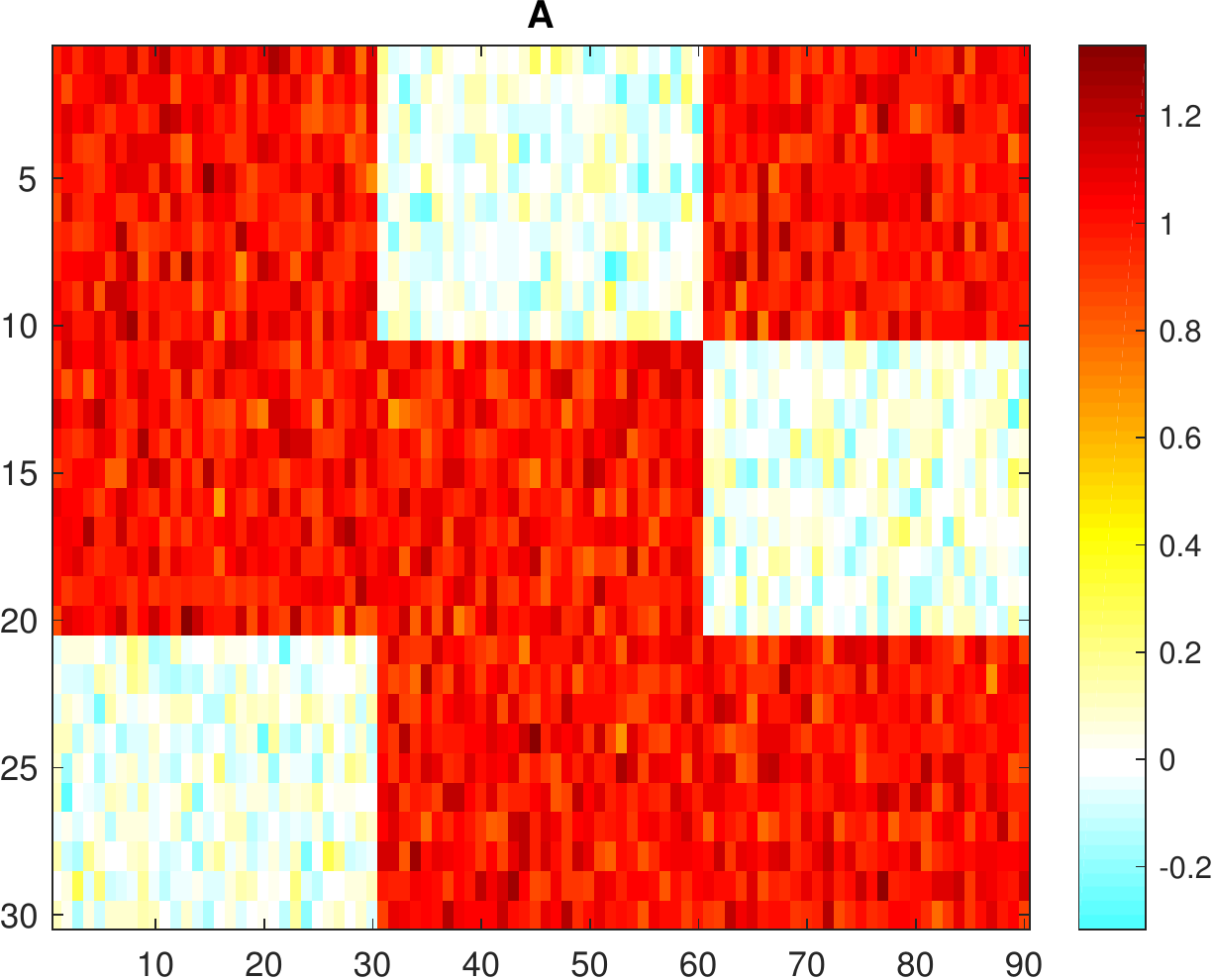}} 
	\scalebox{0.35}{\includegraphics[clip=true, trim=0in 0in 0in 0.15in]{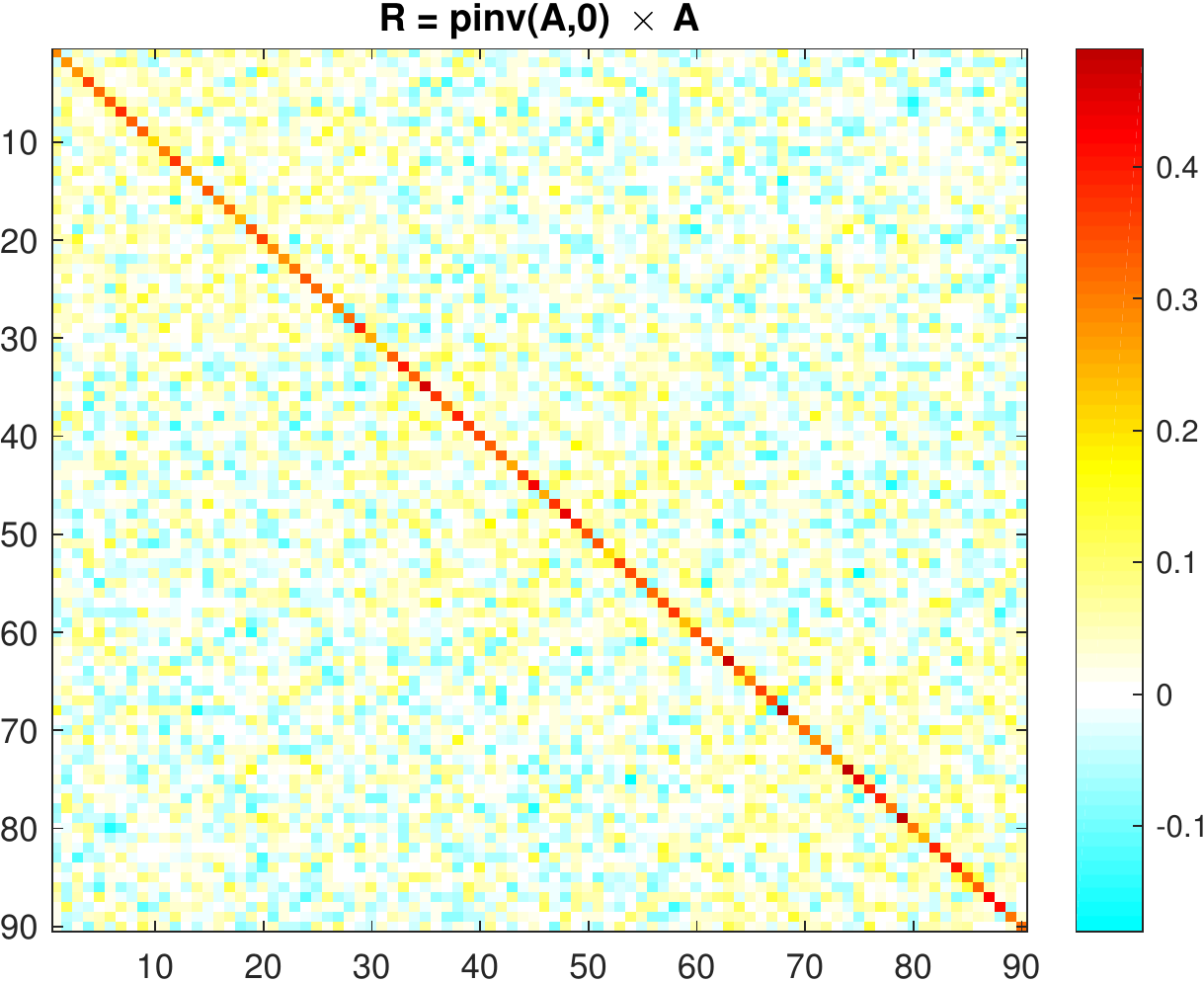}} 
	\scalebox{0.35}{\includegraphics[clip=true, trim=0in 0in 0in 0.15in]{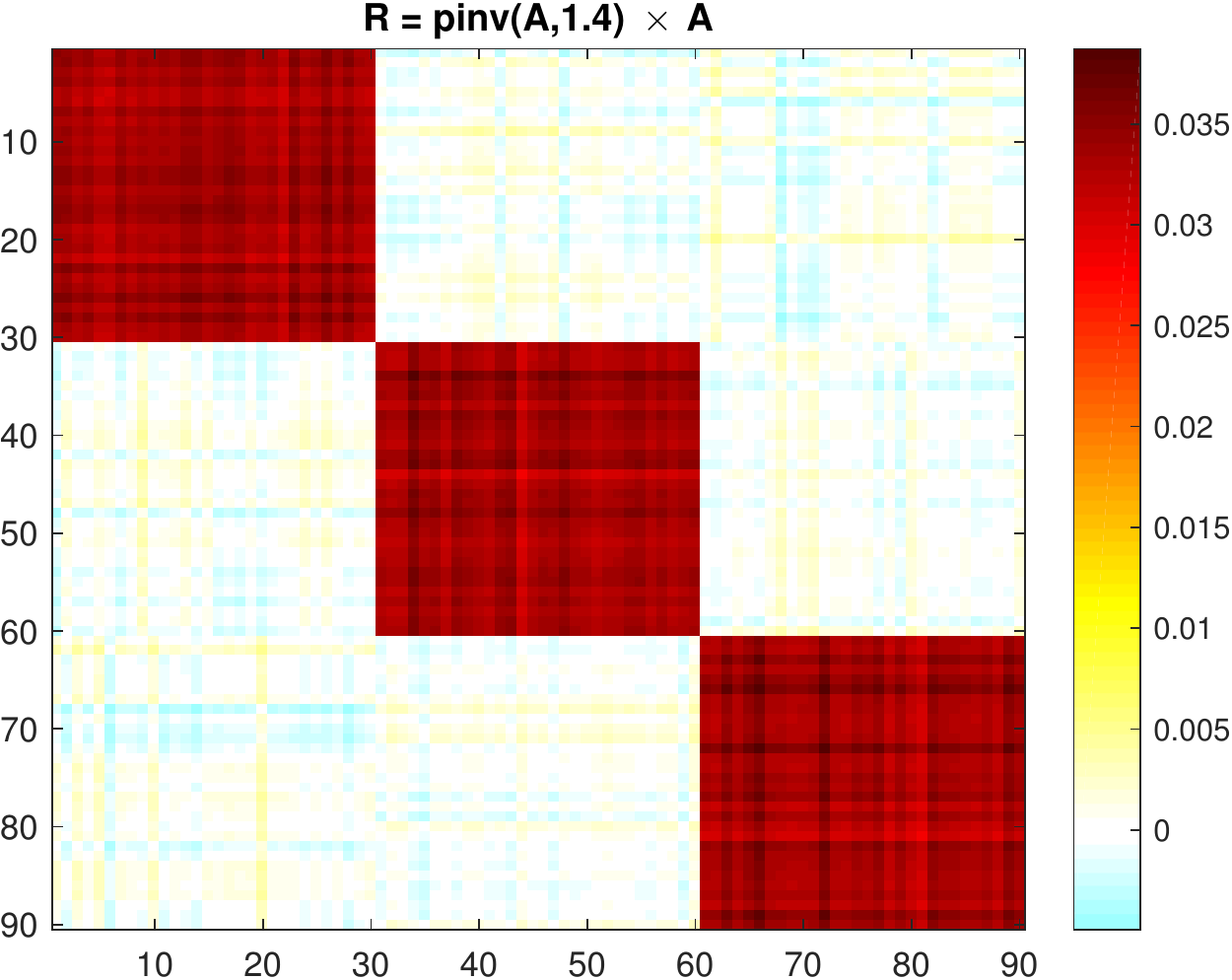}} 
	\caption{Noisy version of data matrix (left), resolution matrix estimate with no regularization (middle), and with regularization (right), demonstrating false gains in resolution due to noise and their correction by regularization.}
	\label{fig_noisynetwork}
\end{figure}

Thus far we have neglected the role of noise and other sources of error. 
The typical inverse problems approach to noise is to treat it as unwanted variation in the measured output data, which in the model of Eq. (\ref{eq_inverse_nbhd}) would imply the following,
\begin{align}
\bb A \bb x_k  + \bb n_k = \bb a_k,
\label{eq_inverse_nbhd_noisy}
\end{align}
where $\bb n_k$ is a random vector of noise signal.
In inverse problems, such noise effects are known to cause a loss of resolution
\cite{dillon_computational_2016}.
In our adaptation of this perspective to connectivity estimation, however, we must take additional care; any signal noise in our measurements $\bb a_k$ will also afflict our forward model $\bb A$, as $\bb a_k$ is itself a column from $\bb A$. 
E.g. we have $\bb A_{\bb N} = \bb A + \bb N$, where $\bb N$ is a matrix of random noise signals.
In other words, signal noise directly becomes model error as well. 
This is especially problematic because it has an opposite effect, making the resolution appear to be higher than it truly is. 
Consider the case where the true $\bb A$ should have low rank and be singular. 
The addition of white noise would result in an invertible (though perhaps poorly-conditioned) $\bb A_{\bb N}$,  which would appear to have maximum resolution, yielding the identity matrix from $\bb R = \bb A_{\bb N}^\dagger \bb A_{\bb N} = \bb A_{\bb N}^{-1} \bb A_{\bb N} = \bb I$.
This is demonstrated in Fig. \ref{fig_noisynetwork}, where a noisy version of the $\bb A$ matrix from Fig. \ref{pic_network} results in a resolution matrix that is approimately the identity.
%

Noise is commonly addressed in inverse problems by regularization of the resolution matrix using Tikhonov regularization \cite{an_simple_2012}.
This effectively replaces Eq. (\ref{eq_nbhd_opt}) with the following regularized regression problem,
\begin{align}
\begin{array}{c}
\bb x_k^* = \arg\underset{\bb x_k}{\min} \; \Vert \bb A \bb x_k  - \bb a_k \Vert_2^2 + \mu \Vert \bb x_k \Vert_2^2,
\end{array}
\label{eq_nbhd_opt_l2}
\end{align}
which requires choice of a regularization parameter $\mu$.
In the underdetermined case (i.e., more voxels than time samples) this has analytical solution,
\begin{align}
\bb x_k^* = \bb A^T (\bb A \bb A^T + \mu \bb I)^{-1}\bb a_k,
\label{eq_xk_mu}
\end{align}
defining the $\ell_2$-regularized pseudoinverse,
\begin{align}
\bb A^\dagger_\mu = \bb A^T (\bb A \bb A^T + \mu \bb I)^{-1},
\label{pinv_mu}
\end{align}
and corresponding $\ell_2$-regularized resolution matrix 
\begin{align}
\bb R_\mu = \bb A^\dagger_\mu \bb A.
\label{eq_Rmu}
\end{align}
This is the approach taken in \cite{dillon_regularized_2017}.
Note that a regularized resolution matrix is the product of the noisy dataset with a regularized version of its pseudoinverse.

A difficulty of regularization methods is of course the need to choose a regularization parameter. 
However, we note that such decisions are already routinely made in fMRI preprocessing.
In particular, a common practice is to truncate the singular value decomposition (SVD) of the data, in order to remove weak components \cite{caballero-gaudes_methods_2017}. 
Further, a number of researchers have developed techniques and guidelines for selection of the proper cutoff for such preprocessing steps \cite{strother_evaluating_2006, churchill_optimizing_2012,yourganov_dimensionality_2011}. 
Next we will show that if a dataset has been preprocessed in this way, then the resulting resolution matrix can be viewed as having been regularized.  

Consider the $m \times n$ dataset $\bb A$ where $m<n$ and $\bb A$ is full row rank. 
The SVD is
\begin{align}
\bb A = \bb U \bb S \bb V^T = \sum_{i=1}^m \sigma_i \bb u_i \bb v_i^T
\label{eq_A_SVD}
\end{align}
where $\bb U$ and $\bb V$ are left and right singular vectors with columns $\bb u_i$ and $\bb v_i$, respectively, and $\bb S$ is a $m \times n$ diagonal matrix of singular values $\sigma_i$.
The truncated SVD of $\bb A$ is
\begin{align}
\bb A_r = \bb U_r \bb S_r \bb V_r^T = \sum_{i=1}^r \sigma_i \bb u_i \bb v_i^T,
\label{eq_Ar}
\end{align}
where $r$ is the rank to which the SVD is truncated, $\bb U_r$ and $\bb V_r$ are the first $r$ columns of $\bb U$ and $\bb V$, respectively, and $\bb S_r$ is the first $r$ rows of $\bb S$. 
The truncated-SVD (TSVD) regularized solution to Eq. (\ref{eq_inverse_nbhd_noisy}) would then be \cite{hansen_truncatedsvd_1987},
\begin{align}
\hat{\bb x}_k = \bb V_r \bb S_r^{-T} \bb U_r^T \bb a_k = \bb A_r^\dagger \bb a_k,
\label{eq_tsvd_soln}
\end{align}
where $\bb S_r^{-T}$ is the $n \times r$ diagonal matrix of the inverses of the first $r$ singular values.
This method has long been known to produce similar results to those of $\ell_2$ regularization \cite{hansen_truncatedsvd_1987}.
In Eq. (\ref{eq_tsvd_soln}) we have also defined the TSVD-regularized pseudoinverse $\bb A_r^\dagger$ by analogy with Eq. (\ref{pinv_mu}),
\begin{align}
\bb A_r^\dagger = \bb V_r \bb S_r^{-T} \bb U_r^T = \sum_{i=1}^r \sigma_i^{-1} \bb v_i \bb u_i^T.
\end{align}
Note that we use Greek subscripts to denote the $\ell_2$-regularized pseudoinverse, and Latin subscripts to denote the SVD-regularized version.
By computing the resolution matrix using this dimensionality-reduced data set, we get the TSVD-regularized resolution matrix,
\begin{align}
\bb R_r = \bb A^\dagger_r \bb A_r = \bb A^\dagger_r \bb A.
\end{align}
So by leveraging results for optimal choice of preprocessing, in this case for truncating the SVD of the data to eliminate weak components, we get a regularized resolution matrix estimate.

\subsection{Spatial smoothing}

%
%
%
%
%
%
%
%
%

%
%
%

Spatial smoothing is another common preprocessing step used for fMRI data \cite{strother_evaluating_2006}, which also can be viewed as a regularization of the resolution matrix. 
In this case, the effect is immediately visible since the resolution matrix itself takes on the smothing operation, as a ``blurring'' of the identity.
Earlier, with Fig. \ref{img_identity_vs_reso}, we noted the interpretation of $\bb R$ as a blurring operator \cite{backus_resolving_1968}. 
Consider that the least-squares solution to Eq. (\ref{eq_inverse_nbhd}) is  
\begin{align}
\hat{\bb x}_k &= \bb A^\dagger \bb A \bb x_k \notag \\
&= \bb R \bb x_k.
\label{eq_a_hat}
\end{align}
So the least-squares solution $\hat{\bb x}_k$ is a ``blurred" version of the true solution, with blurring described by the resolution matrix.
In particular, note that this blurring operator projects the true solution onto the rowspace of $\bb A$.
Such projections are known as estimable functions \cite{rodgers_estimable_1982}, meaning features which may be estimated even when the forward model is not invertible \cite{dillon_robust_2017}.
In computational imaging, resolution cells may be viewed as local averages which may be estimated from the data.
Similarly in a connectivity estimation problem, while we may not be able to estimate the edge weight for a particular edge to a node, we may be able to estimate the average weight over multiple edges connecting to the node.
The averaging operator is the corresponding row of the resolution matrix.

\begin{figure}[h!] \centering 
	\scalebox{0.6}{\includegraphics[clip=true, trim=0in 0in 0in 0in]{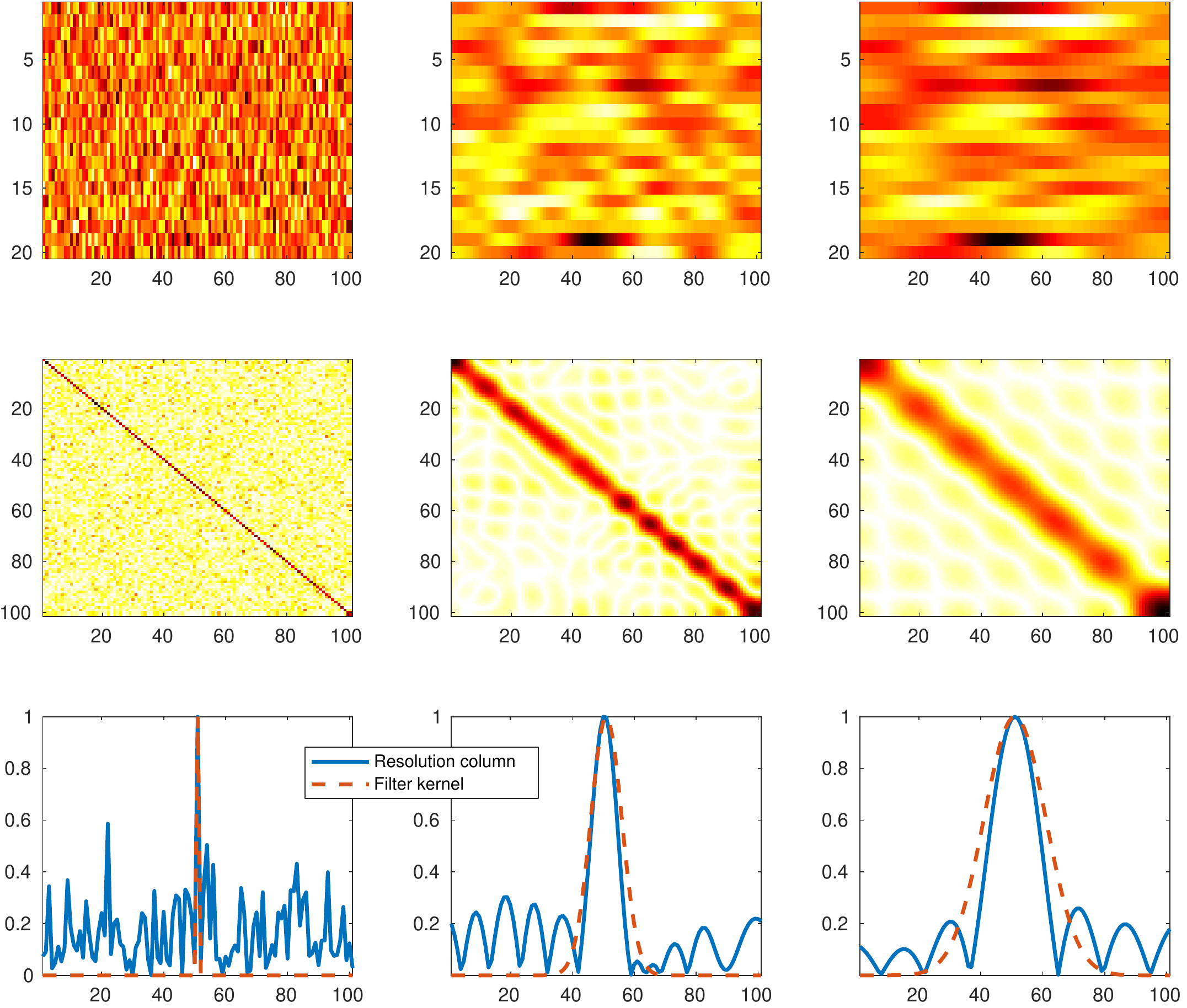}} 
	\caption{Noisy version of data matrix (left), resolution matrix estimate with no regularization (middle), and with regularization (right), demonstrating false gains in resolution due to noise and their correction by regularization.}
	\label{fig_noise_filt_sim}
\end{figure}

The spatial smoothing kernel, therefore, directly relates to the resulting blurring kernel of the resolution matrix.
Spatial smoothing creates local correlations between nearby time-series, which the resolution matrix subsequently describes.
This is demonstrated in the simulation of Fig. \ref{fig_noise_filt_sim}.
In this case the resolution matrix is an approximate identity, while the spatial smoothing produces a similarly-smoothed version of the identity matrix.
We also see that the blurring kernel described by the resolution matrix columns is very similar to the smoothing kernel used on the data.
This produces an important regularization effect in clustering techniques.

\subsection{Resolution Clustering}

The example in Fig. \ref{pic_network} motivates the idea of separating the sub-networks by clustering the resolution matrix. 
And indeed this was demonstrated in \cite{dillon_regularized_2017} using the 
the method is presented in Algorithm 2. 
\begin{algorithm}
	\begin{algorithmic}[1]
		\caption{Resolution Clustering}
		\STATE Form standardized data matrix $\bb A$ containing time series as columns. Choose regularization parameter $\mu$ and number of clusters $K$. 
		\STATE Compute $\bb A^{\dagger}_\mu$, the regularized pseudoinverse of $\bb A$.
		\STATE Apply $k$-means clustering to the columns of $\bb R = \bb A^{\dagger}_\mu \bb A$ with $K$ clusters. 
		\label{pinv_algo}
	\end{algorithmic}
\end{algorithm}
%
%
%
%
%
%
%
%

In \cite{dillon_regularized_2017} a memory-efficient clustering algorithm was also provided to handle large datasets $\bb A$ for which $\bb R = \bb A^{\dagger}_\mu \bb A$ would be too large to store in memory.
For example, the Philadelphia Neurodevelopmental Cohort fMRI data volumes are $79 \times 95 \times 79$ voxels, for  124 time samples, resulting in a $\bb A$ matrix of size $592895 \times 124$. 
The resulting resolution matrix would be $592895 \times 592895$ (as is the covariance matrix for forming network affinity matrices). 
However we never need store this in its entirety. 

We will describe that method here, as the approach can be used for clustering regularized resolution matrices as well as for clustering the sample covariance matrix via $\bb A^T \bb A$.
A basic $k$-means algorithm is presented in Algorithm 3. 
\begin{algorithm}
	\begin{algorithmic}[1]
		\caption{$k$-means applied to data columns of matrix $\bb M$}
		\STATE Choose number of clusters $K$ and initialize cluster centers $\bb c_k, \; k=1,...,K$
		\WHILE {Convergence criterion not met}
		\STATE Label each column with that of nearest cluster center: $l_k = \arg \min_i D_{ik}$, where $D_{ik}$ is distance between column $\bb m_k$ and cluster center $\bb c_i$
		\STATE Recalculate cluster centers as mean over data columns with same label: $\bb c_i = \frac{1}{\vert S_i \vert}\sum_{j \in S_i} \bb m_j$,  where $S_i = \left\{k | l_k = i \right\} $. 
		\ENDWHILE
		\label{kmeans_algo}
	\end{algorithmic}
\end{algorithm}
Here we consider the general case of clustering columns of $\bb M = \bb P \bb Q$, where $\bb P$ (of size $n \times m$) and $\bb Q$ (of size $m \times n$) are stored in memory but $\bb M$ is too large to store.
So for resolution $\bb M = \bb R$ and $\bb P$ and $\bb Q$ are $\bb A^\dagger$ and $\bb A$, respectively.
First note that the squared distances between a given center $\bb c_i$ and a column $\bb m_k$ of $\bb M$ can be calculated as
\begin{align}
D_{ik}^2 &= \Vert \bb c_i - \bb m_k \Vert_2^2 \notag \\
&= \bb c_i^T \bb c_i + \bb m_k^T \bb m_k - 2 \bb c_i^T \bb m_k \notag \\
&= \bb c_i^T \bb c_i + \bb m_k^T \bb m_k - 2 \bb c_i^T \bb P \bb q_k.
\label{eq_Dik2}
\end{align}
Since we are only concerned with the class index $i$ of the cluster with the minimum distance to each column, we do not need to compute the $\bb m_k^T \bb m_k$ term. So we can compute
\begin{align}
l_k &= \arg \min_i D_{ik}^2 \notag \\
&= \arg \min_i \left\{ \bb c_i^T\bb c_i - 2 \bb c_i^T \bb P \bb q_k \right\}.
\label{eq_lk}
\end{align}
By forming a matrix $\bb C$ with cluster centers $\bb c_i$ as columns, we can efficiently compute the cross term in brackets for all $i$ and $k$ as $(\bb C^T \bb P) \bb Q$, a $K$ by $n$ matrix.

Along similar lines, we can efficiently compute the mean over columns in each cluster by noting that the mean over a set $S$ of columns can be written as
\begin{align}
\bb c_i &= \frac{1}{\vert S \vert}\sum_{j \in S} \bb m_j 
= \frac{1}{\vert S \vert}\sum_{j \in S} \bb P \bb q_j 
= \frac{1}{\vert S \vert} \bb P \sum_{j \in S} \bb q_j.
\label{eq_ci}
\end{align}
So for clustering the resolution matrix, we have that clustering of $\bb R$ requires additional storage for a matrix the same size as $\bb A$ (i.e., $\bb A^\dagger$), and roughly double the number of calculations as conventional clustering, with two matrix-vector multiplies replacing each single one in the conventional algorithm.

%
%

%

%
%
%
%

\subsection{Spectral Resolution Clustering}

For TSVD regularization, we can use Algorithm 2 
and the memory-efficient technique of the previous section, replacing $\bb A_\mu^\dagger$ with $\bb A_r^\dagger$. 
However we can find an potentially even more efficient algorithm, which can utilize preprocessing calculations that are already being performed. 
First note that using the SVD of $\bb A = \bb U \bb S \bb V^T$, we get
\begin{align}
\bb R = \bb A^\dagger \bb A = \bb V \bb V^T.
\end{align}
Similarly for the truncated SVD, $\bb A_r = \bb U_r \bb S_r \bb V_r$, we get
\begin{align}
\bb R_r = \bb A_r^\dagger \bb A = \bb A_r^\dagger \bb A_r = \bb V_r \bb V_r^T. 
\label{eq_Rr_svd}
\end{align}
Using the result from the previous section, now with $\bb P = \bb V_r$ and $\bb Q =  \bb V_r^T$, we have from Eq. (\ref{eq_ci}),
\begin{align}
\bb c_i &= \frac{1}{\vert S \vert} \bb P \sum_{j \in S} \bb q_j \notag \\ 
        &= \frac{1}{\vert S \vert} \bb V_r \sum_{j \in S} \bb v_r^{(j)}, 
\label{eq_ci_VVT}
\end{align}
where $\bb v_r^{(j)}$ is the (transposed) $j$th row of $\bb V_r$.
%
%
Then combining Eq. (\ref{eq_ci_VVT}) and Eq. (\ref{eq_Dik2}), we get
\begin{align}
D_{ik}^2 &= \Vert \bb c_i - \bb m_k \Vert_2^2 \notag \\
&= \left\Vert \frac{1}{\vert S \vert} \bb V_r \sum_{j \in S} \bb v_r^{(j)} - \bb V_r \bb v_r^{(k)} \right\Vert_2^2 \notag \\
&= \left\Vert \frac{1}{\vert S \vert} \sum_{j \in S} \bb v_r^{(j)} - \bb v_r^{(k)} \right\Vert_2^2 
\label{eq_Dik2}
\end{align}
So the distances are equal to the distances between rows of $\bb V_r$ and centers resulting from the average over the rows of $\bb V_r$ corresponding to the cluster.
The method is summarized in Algorithm 4.
\begin{algorithm}
	\begin{algorithmic}[1]
		\caption{TSVD-regularized Resolution Clustering.}
		\STATE Form standardized data matrix $\bb A$ containing time series as columns. Choose rank truncation $r$ and number of clusters $K$. 
		\STATE Compute $r$ singular vectors  $\bb (\bb v_1, ..., \bb v_r) = \bb V_r$ corresponding to largest $r$ singular values of $\bb A$.
		\STATE Apply $k$-means clustering to the rows of $\bb V_r$ with $K$ clusters. 
		\label{TSVD_spectral_reso_algo}
	\end{algorithmic}
\end{algorithm}

Next consider that in typical approaches to spectral clustering start by forming the graph from $\bb A$, for example by computing pairwise correlations between columns (representing time series) then zeroing values below a threshold or otherwise computing a distance metric that yields a non-negative adjacency matrix. 
If we stopped at the raw correlation estimate, and did not perform these heuristic adjustments, we would compute a scaled version as $\bb A^T \bb A$, which can be viewed as a dense signed adjacency matrix itself.
The eigenvalue decomposition of this matrix gives
\begin{align}
\bb A^T \bb A = \bb V \boldsymbol{\Lambda} \bb V^T, 
\end{align}
where $\boldsymbol{\Lambda}$ is a diagonal matrix of eigenvalues. 
So Algorithm 4 can be viewed as a close relative of Algorithm 1 applied to this version of an adjacency matrix (recall that \cite{meila_learning_2002} showed that one can equivalently use eigenvectors of a version of the adjacency matrix).
Hence our principled goal of segmenting network resolution yields a variation on spectral clustering.
We also have a direct interpretation of the choice of model order (i.e., the truncation rank $r$) based on regularization of the neighborhood estimation problem for predicting node activity. 
And further, this choice may be considered separately from the choice of number of clusters. 
Next we show how to relate the $\ell_2$-regularization method  into this same framework.







If we input the SVD of $\bb A$ from Eq. (\ref{eq_A_SVD}) into Eq. (\ref{pinv_mu}), we get \cite{hansen_truncatedsvd_1987}
\begin{align}
\bb A^\dagger_\mu 
 &= \sum_{i=1}^m \frac{\sigma_i}{\sigma_i^2 + \mu} \bb v_i \bb u_i^T. 
\label{pinv_mu_svd}
\end{align}
Inputting this into the resolution matrix of Eq. (\ref{eq_Rmu}) gives
\begin{align}
\bb R_\mu 
 &= \sum_{i=1}^m \frac{\sigma_i^2}{\sigma_i^2 + \mu} \bb v_i \bb v_i^T 
   = \bb V_\mu \bb V_\mu^T. 
   \label{eq_Rmu_svd}
\end{align}
where we have defined $\bb V_\mu$ as a matrix of weighted singular vectors 
\begin{align}
\bb V_\mu = \bb V \bb D_{\bb w},
\end{align}
using $\bb D_{\bb w}$, a diagonal matrix with diagonal $\bb w$ where $w_i = \sqrt{\frac{\sigma_i^2}{\sigma_i^2 + \mu}}$.
This suggests the general method of Algorithm 5, where have the following options for $\bb w$,
\begin{align}
(\bb w_r)_i &= \begin{cases}
1, i \le r \\
0, i > r
\end{cases} \\
(\bb w_\mu)_i &= \sqrt{\frac{\sigma_i^2}{\sigma_i^2 + \mu}}.
\end{align}
\begin{algorithm}
	\begin{algorithmic}[1]
		\caption{Weighted Spectral Resolution Clustering).}
		\STATE Form standardized data matrix $\bb A$ containing time series as columns. Choose rank truncation $r$, number of clusters $K$, and weighting $\bb w$. 
		\STATE Compute $r$ singular vectors  $\bb (\bb v_1, ..., \bb v_r) = \bb V_r$ corresponding to largest $r$ singular values of $\bb A$.
		\STATE Apply $k$-means clustering to the rows of $\bb V \bb D_{\bb w}$ with $K$ clusters, where $\bb D_{\bb w}$ is the diagonal matrix with $\bb w$ on the diagonal. 
		\label{weighted_spectral_reso_algo}
	\end{algorithmic}
\end{algorithm}
This further provides the ability to generalize to other weightings which may yield more optimal estimators in the neighborhood estimation problem.

\section{Real Data Results}

\begin{figure}[h!] \centering 
	\scalebox{0.5}{\includegraphics[clip=true, trim=0in 0in 0in 0in]{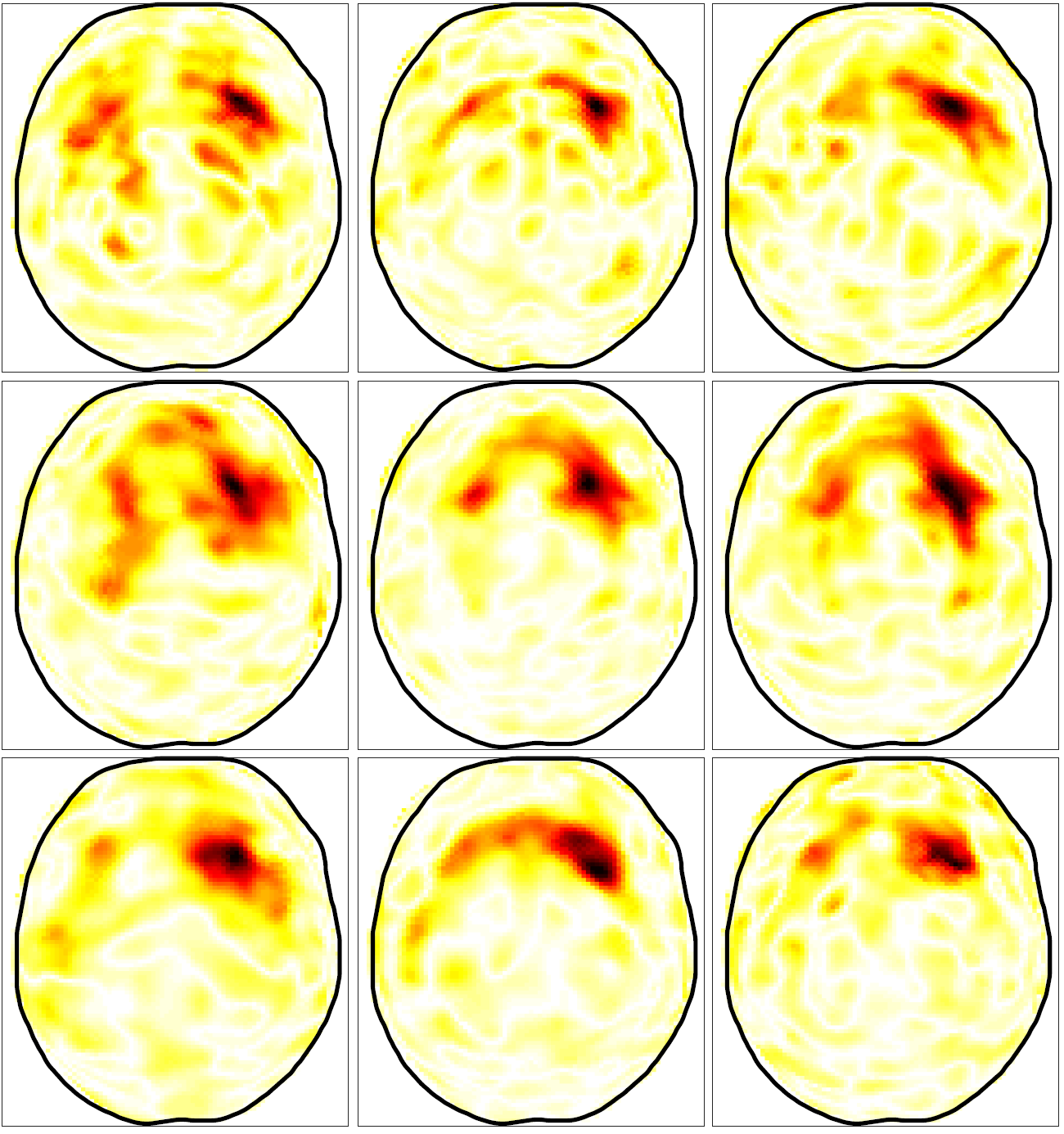}} 
	\caption{Single resolution cell for voxel in right precentral gyrus for individuals from PNC dataset; visualized as top view of brain, with each voxel colored by magnitude of its contribution to the resolution cell; the stronger the signal, the more ``unresolvable'' the given voxel is from the chosen point in precentral gyrus; each column shows the resolution cell for a different scan, and each row represents a different subject. Note that this is a column of the resolution matrix, reformed into a 3D image, then viewed from above.} 
	\label{fig_example_pnc_reso_cells}
\end{figure}

We used data from the Philadelphia Neurodevelopmental Cohort \cite{satterthwaite_philadelphia_2016}, which contains three scans for each subject, a resting-state scan and two task scans. 
The data was preprocessed using SPM, which included spatial smoothing with a 5 mm  kernel, and registration to normalized coordinates.
We then formed masks of the brain region by setting a threshold, and selected the 100 subjects which had the highest common overlap between masks. 
The data was downsampled by a factor of three in each dimension, which allowed it to be made small enough to form the adjacency matrix in memory for other spectral clustering methods (as efficient code for most methods was not available).
Examples of resolution estimates for this dataset are provided in Fig. \ref{fig_example_pnc_reso_cells}.

%
%
%
%

\subsection{Hyperparameter Estimation}

We used a cross-validation method to determine the parameters for the TSVD and $\ell_2$-regularized methods, based on the prediction of activity.   
Each choice of parameter was evaluated by the following steps, for each subject:
\begin{enumerate}
	\item Separate data into training and test sets, forming separate matrices $\bb A^{(train)}$ and $\bb A^{(test)}$; standardize and preprocess independently (to the degree possible).
	\item Using training set data, compute regularized predictor $\bb x_k^{(train)}$ for every voxel $k$, via Eq. (\ref{eq_xk_mu}) or (\ref{eq_tsvd_soln}), depending on method.
	\item Set predictor values in vicinity of voxel $k$ to zero in order to eliminate self-loops.
	\item Rescale predictor to compensate for exclusion of vicinity of voxel $k$.
	\item Using test set data, compute residual $\Vert\bb A^{(test)} \bb x_k^{(train)} - \bb a_k^{(test)}\Vert$ for every voxel $k$.
	\item Average residual over all voxels and all cross-validation folds, and choose parameter which minimizes the residual.
\end{enumerate}
For the training and test sets, we tried splitting single scans into three parts, by separating the time series into thirds.
We also tried using one of a given subject's scans as training set and testing against the other two. 
The latter method has the advantage that the training and test sets are preprocessed completely independently (i.e., normalization of coordinates and temporal filtering) so there is no opportunity for sharing information.
We found the optimals to be roughly the same for either approach to cross-validation, so provide the results for the multiple scan approach.
We used a similar procedure to test different choices of temporal filtering, but the applying filtering resulted in only a small variation in prediction accuracy, so we omitted this preprocessing step.

Recall that $\bb x_k$ can be viewed as a column of a weighted adjacency matrix $\bb X$.
So in effect we are using $\bb A^{(train)}$ to estimate connectivity, then using this network to predict the activity on $\bb A^{(test)}$.
It is necessary to exclude the region around the $k$th voxel because this would include a self-loop, which makes predictions that are trivially accurate. 
To correct for this exclusion of part of the predictor, we estimate an optimal scalar for the predictor (also using the training set) by finding the optimal value to minimize
\begin{align}
\alpha^* = \arg \min_\alpha  \sum_k\Vert\bb A^{(train)} \bb x_k^{(train)}\times \alpha - \bb a_k^{(train)}\Vert^2
\label{eq_optimal_alpha}
\end{align}
which has analytical solution,
\begin{align}
\alpha^* = \frac
{\sum_{i,j} A^{(train)}_{i,j}\left(\bb A^{(train)} \bb X^{(train)}\right)_{i,j}}
{\sum_{i,j} \left(\bb A^{(train)} \bb X^{(train)}\right)_{i,j}^2}.
\end{align}
We computed prediction residual with and without this scalar adjustment.

\begin{figure}[h!] \centering 
	\scalebox{0.6}{\includegraphics[clip=true, trim=0in 0in 0in 0in]{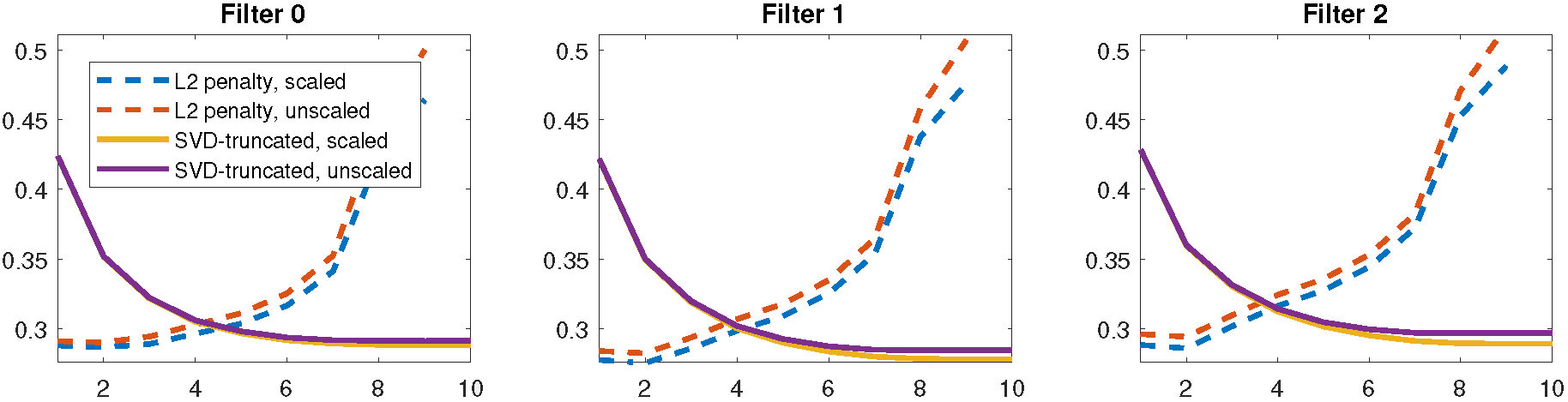}} 
	\scalebox{0.6}{\includegraphics[clip=true, trim=0in 0in 0in 0in]{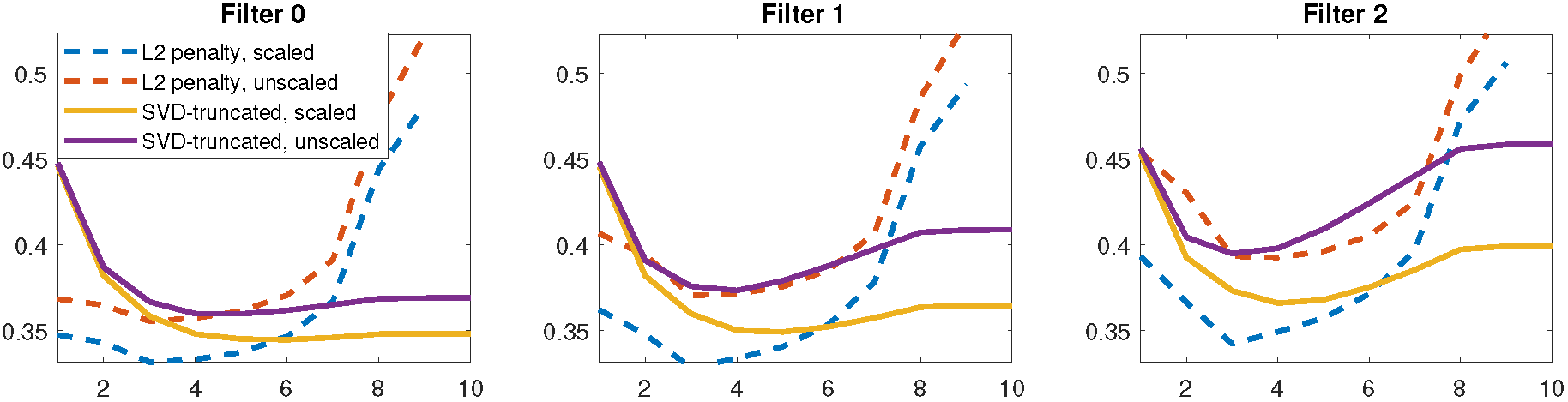}} 
	\scalebox{0.6}{\includegraphics[clip=true, trim=0in 0in 0in 0in]{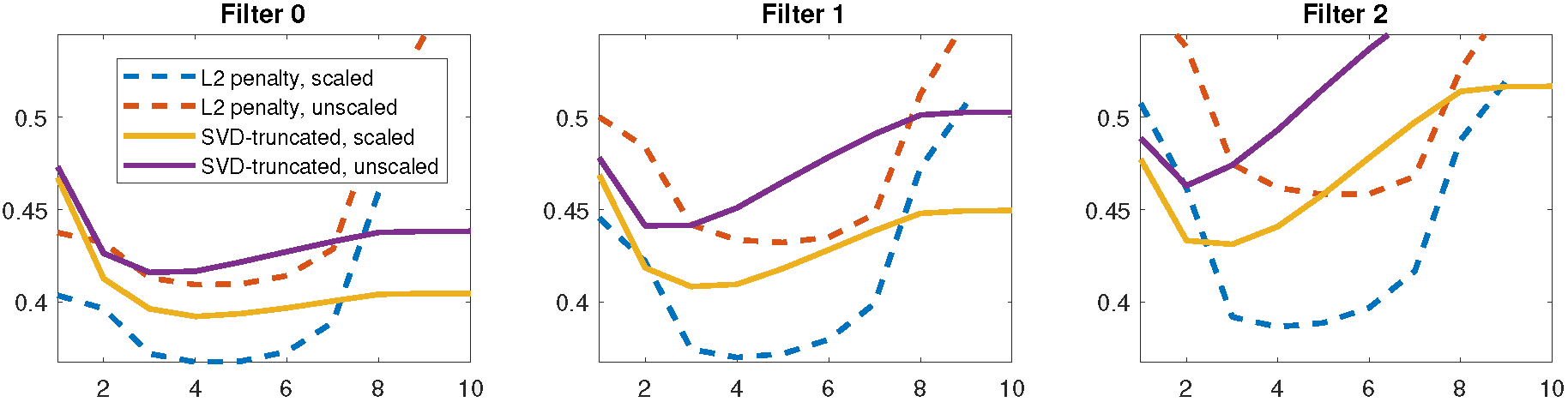}} 
	\caption{Average 3-fold cross-validated fractional residual plotted versus regularization parameter (10 choices). Excluded region size = 5 mm (top), 10 mm (middle), 15 mm (bottom). Spatial smoothing kernel size = 5 mm (left column), 10 mm (middle column), 15 mm (right column). Plots are given both with and without the $\alpha^*$ scaling. Optimal of 3-5 for SVD corresponds to 30-50 percent singular vectors retained. Optimal of 3-5 for L2 corresponds to $0.1 \sigma_{max}$ to  $0.3 \sigma_{max}$.}
	\label{fig_cv_svd_vs_rr}
\end{figure}

Figure \ref{fig_cv_svd_vs_rr} gives the results of the three-fold cross-validation test for ten different choices of regularization for both the $\ell_2$ and TSVD methods. For the TSVD, the ten steps correspond to tenths of the total number of nonzero singular values. So the minimum around 3 to 5 on the horizontal axis corresponds to 30 to 50 percent dimensionality reduction for the TSVD method. 
For the $\ell_2$ regularization we used the following multiples of the maximum singular value: 0.001, 0.01, 0.1, 0.2, 0.3, 0.5, 1.0, 5.0, 10.0. 
So the minimum around 3 to 5 corresponds to $0.1 \sigma_{max}$ to  $0.3 \sigma_{max}$.
This range of tests was repeated for multiple choices of spatial smoothing and multiple choices of the exclusion region. 
We see that if the exclusion region is small (5mm) then a choice of no regularization was optimal, which suggests self-loops were still having an effect due to spatial smoothing. 
Otherwise we had relatively consistent results with the $\ell_2$-regularization outperforming the TSVD method, and minima in the 3 to 5 range. 

\subsection{Parcel Comparisons}

%
%
%
%
%
%
%
%
%

\begin{figure}[h!] \centering 
	\scalebox{0.75}{\includegraphics[clip=true, trim=0in 0in 0in 0in]{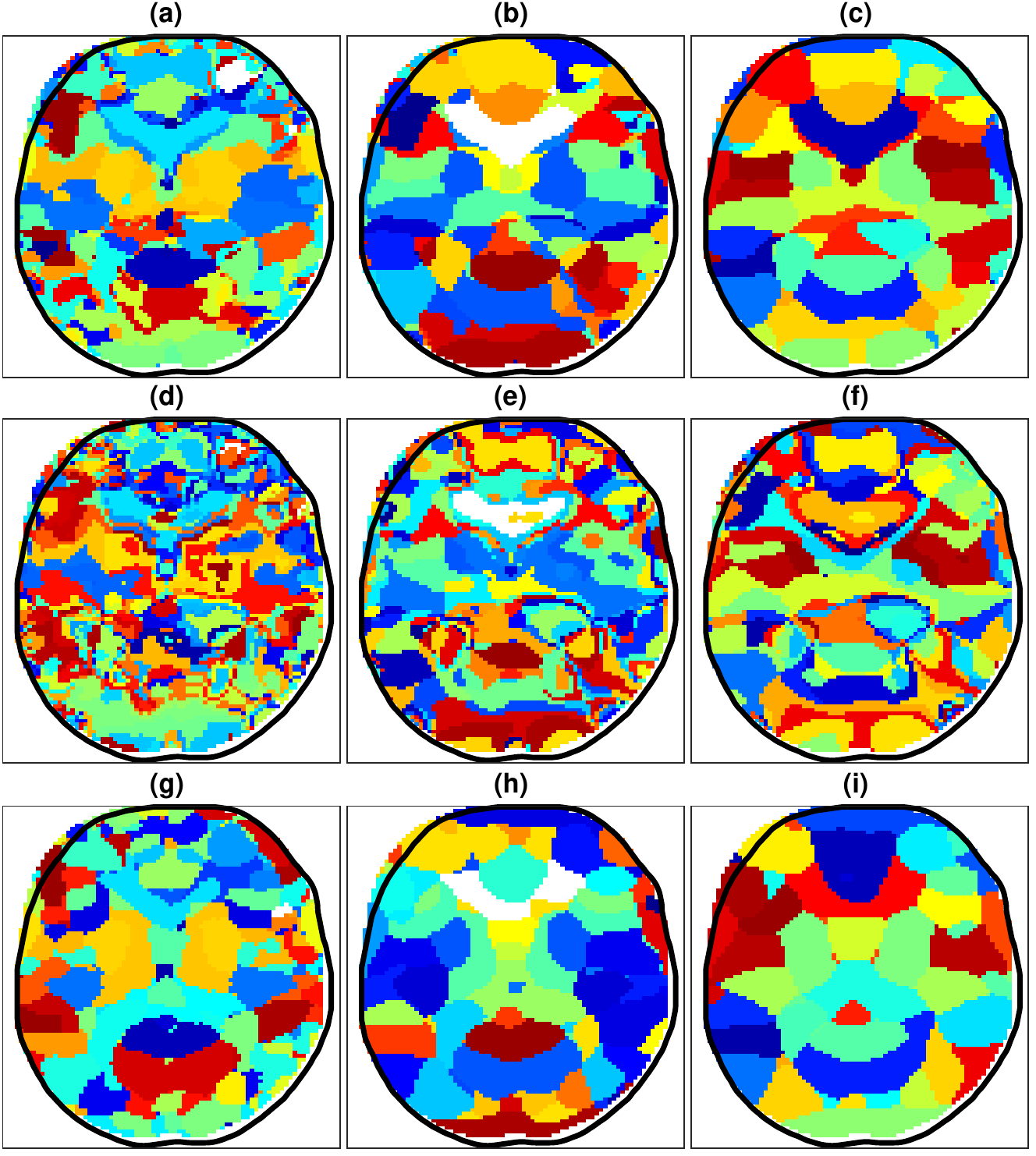}} 
	\caption{Horizontal slice of parcellation for single subject, for three different methods (rows) and three different degrees of spatial smoothing (columns). Top row is the ``A'' method; middle row  shows the ``AA'' method; bottom row gives the ``Rr'' method; left column is with low filtering (5 mm kernel); middle column is medium filtering (10 mm kernel); right column is high filtering (15 mm kernel).}
	\label{fig_example_parcels}
\end{figure}

Next we computed individual parcellations for each of thee three scans for the 100 subjects.
We used the number of clusters as $k=116$ (so we could compare to the predefined parcellation), and used random starting clusters.  
The methods used are listed below:
\begin{description}[font=\sffamily\bfseries, leftmargin=1.5cm, style=nextline]
	\item[Rr] $k$-means clustering of  columns of $\bb R_r$. I.e., truncated-SVD clustering method of Algorithm 4, with cutoff of 40 percent singular values.
	\item[Rl] $k$-means clustering of  columns of $\bb R_\mu$. I.e., 
	$\ell_2$-regularized clustering method of Algorithm 2, with $\mu = 0.3 \sigma_{max}$.
	\item[A] $k$-means clustering of the time-series corresponding to individual voxels. I.e. clustering of (standardized) columns of the matrix $\bb A$.
	\item[Ar] $k$-means clustering of columns of the dimensionality-reduced matrix $\bb A_r$ of Eq. (\ref{eq_Ar}), with a cutoff of 40 percent singular values.	
	\item[AA] $k$-means clustering of columns of the scaled sample covariance matrix $\bb A^T\bb A$.	
	\item[WNN] Spatially-constrained spectral clustering; weighted adjacency matrix formed by pairwise voxel correlations greater than 0.4 for nearest neighbors  (adjacent voxels) only.	
	\item[NN] Spectral clustering of binary graph formed by connecting nearest neighbors (adjacent voxels) only; imaging data was not used.
	\item[XYZ] $k$-means clustering of $3 \times n$ position matrix; each column is the three-dimensional location of a voxel.
	\item[AAL] The 116 Automated Anatomical Labeling regions of interest \cite{tzourio-mazoyer_automated_2002}.
	\item[RNG] Parcels defined by randomly labeling each voxel.
\end{description} 
We tried the standard approaches to spectral clustering but were not able to get reasonable-looking parcellations with them, even with high levels of spatial smoothing. 
Figure \ref{fig_example_parcels} shows an example of the parcellations for a single subject for three of the methods. 

\begin{figure}[h!] \centering 
	\scalebox{0.5}{\includegraphics[clip=true, trim=0in 0in 0in 0in]{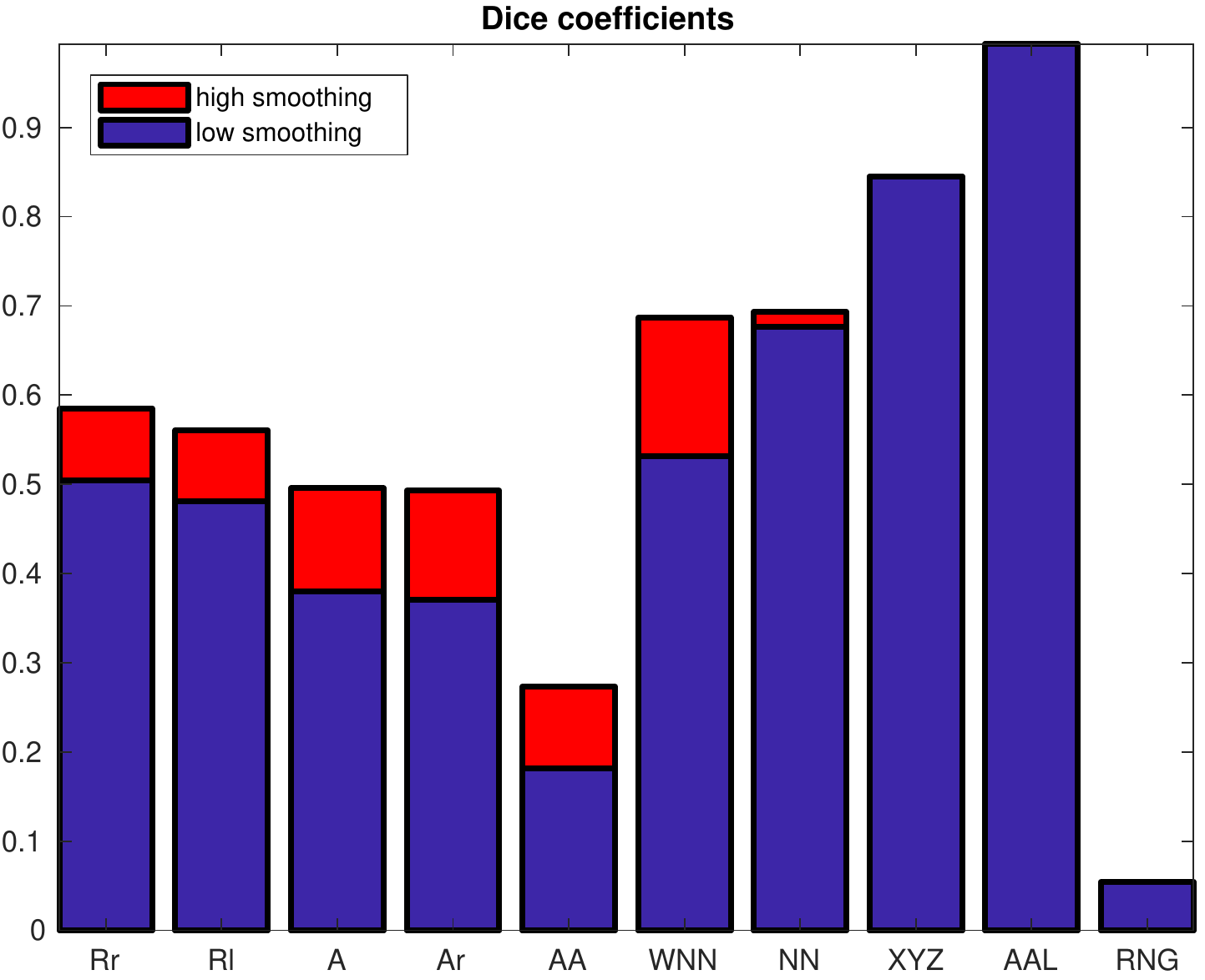}} 
	\caption{Average dice coefficients comparing parcellations between different scans of same subject. Spatial smoothing improves the dice coefficients for the data-dependent methods due to effectively imposing distance regularization.}
	\label{fig_dice}
\end{figure}

First we tested the consistency of results from using different scans for the same subject by computing the average over dice coefficients between a cluster in one scan and the most-similar cluster in the comparison scan. 
The results are given in Fig.  \ref{fig_dice}, for both low (5 mm kernel) and high (15 mm kernel) spatial smoothing.
The averages were rather low, which may result from several causes. 
Notably, the scans were of different types, resting-state versus different task scans, which can account for roughly 30 percent of the variance in the time-series \cite{wang_parcellating_2015}. 
Further, while we chose the subjects with the best-aligned scans, they were still not perfectly aligned.
Consider that the XYZ and NN methods essentially produce tilings of the volume which should be approximately equal for the scans depending how well-aligned they are.
Hence these may serve as something of an upper limit on the data-dependent methods' reproducibility.

We find that the WNN produces the most similar clusters of the data-dependent methods, though at low smoothing its dice coefficients are only slightly higher than the  resolution-based methods. At high smoothing the WNN method appears identical to the NN method. 
We also note that sizable improvements in dice coefficients can be achieved by spatial smoothing, likely due to its effective distance regularization effect, and probably biasing results towards the XYZ and NN methods.

\begin{figure}[h!] \centering 
	\scalebox{0.5}{\includegraphics[clip=true, trim=0in 0in 0in 0in]{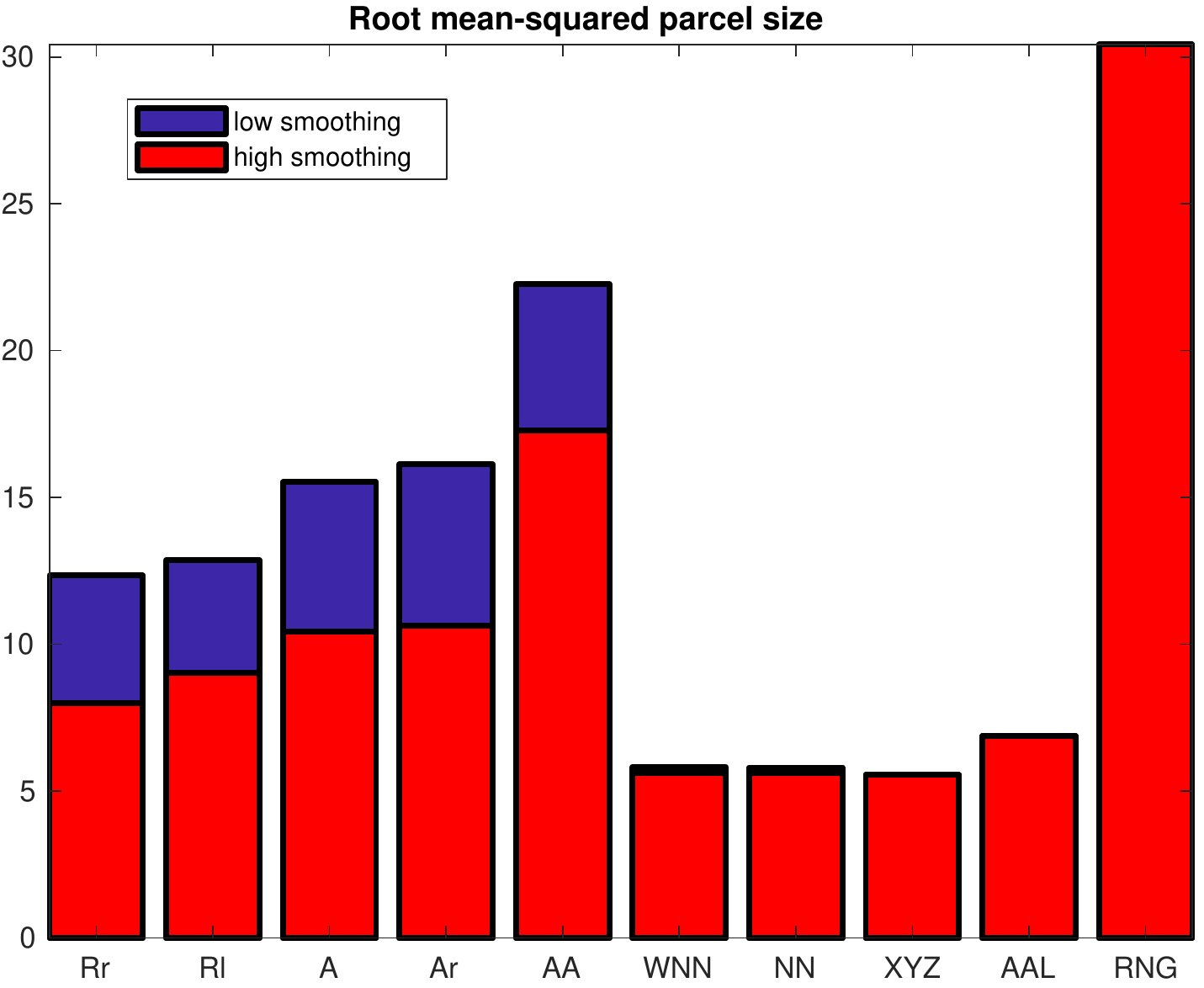}} 
	\caption{Root mean-squared cluster size. Spatial smoothing reduces the parcel size due to effectively imposing distance regularization.}
	\label{fig_r2}
\end{figure}

Fig. \ref{fig_r2} gives a metric of the average parcel size, computed as the square root of the average squared distance between the parcel centroid and each voxel in the parcel.  
Here spatial smoothing reduces the size of parcels for the data-dependent methods, except for WNN. 
As more compact parcels tend to fit more with our prior belief about the modularity of the brain (see Fig. \ref{fig_example_parcels})), we presume smaller parcels here are generally better, at least to some degree. 
Under this assumption, then the best results are achieved by the WNN method, though its immunity to the distance-regularizing effect of spatial smoothing, plus its similarity to the  NN method, makes the validity of the parcel sizes questionable. 
Next-smallest clusters are achieved by the resolution-based methods.

Lastly we computed metrics to test the homogeneity and separation of the parcels.
We tested homogeneity with two metrics; the first was the average unexplained variance within parcels. This is the fraction of remaining signal energy after removing the average signal in the parcel, averaged over parcels. 
Second we computed the average absolute Pearson correlation between every pair of time series within each parcel, excluding self-correlations. 
We tested parcel separation by computing the average absolute correlation between average signals in different parcels.

\begin{figure}[h!] \centering 
	\scalebox{0.5}{\includegraphics[clip=true, trim=0in 0in 0in 0in]{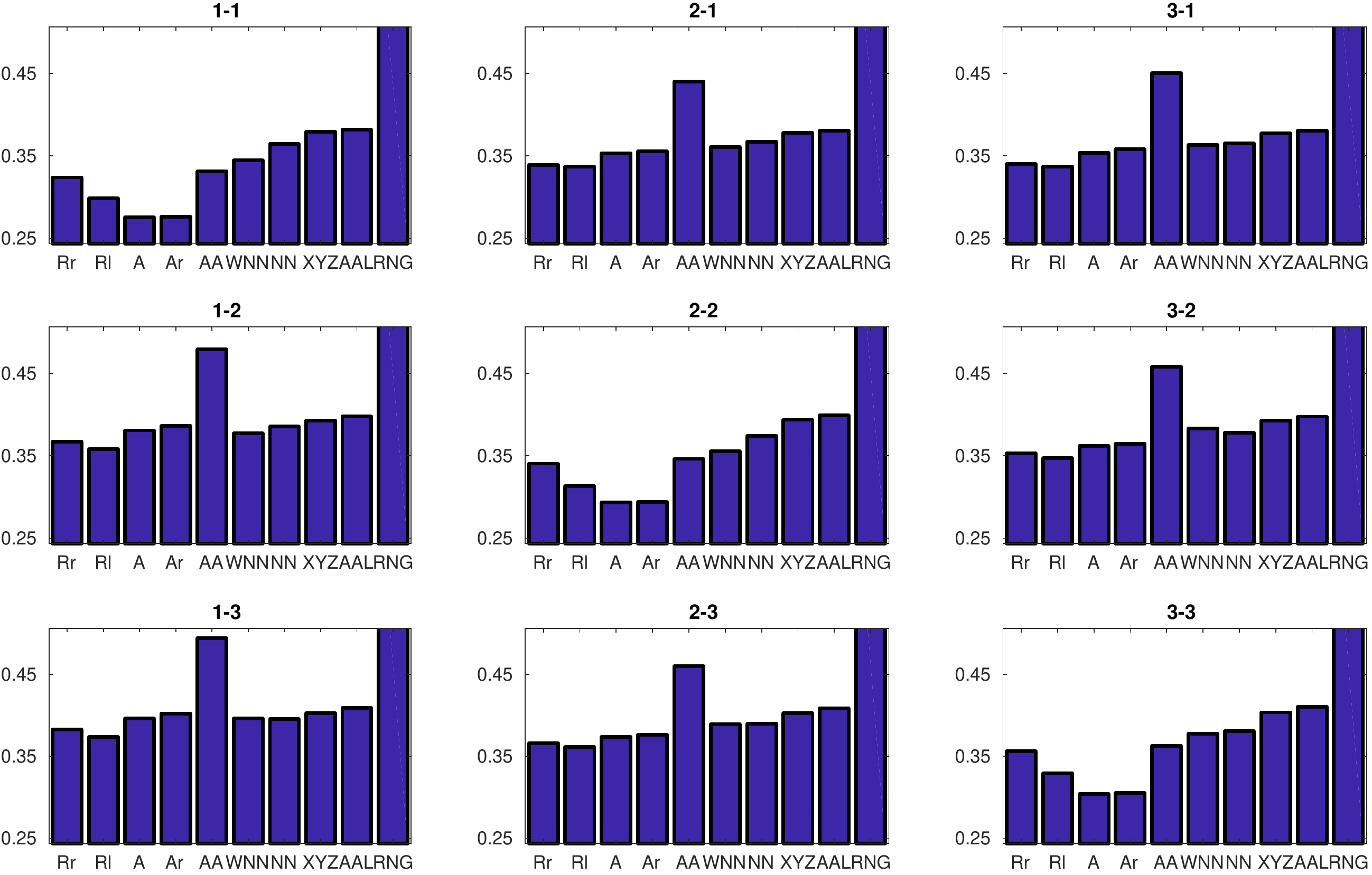}} 
	\caption{Average fractional unexplained variance from applying parcellation from one scan to another scan; figure title $x$-$y$  refers to parcellation scan $x$ and test scan $y$. Note $\bb A$ method is best only for same-scan tests while resolution-based methods are consistently best for cross-scan tests.}
	\label{fig_fuev_x-y}
\end{figure}

Our results agreed with those reported elsewhere \cite{arslan_human_2018}, where it was noted that despite the superior repeatability (i.e. higher dice coefficients) of spectral clustering, the homogeneity was inferior to that of $k$-means clustering (which we are calling the ``A" method).
This makes sense as $k$-means is explicitly a greedy algorithm for optimizing this metric. 
However, we further tested the validity of these results by calculating the metrics for parcels generated using one scan, when applied to another scan for the same subject. 
Fig. \ref{fig_fuev_x-y} gives the fractional unexplained variance for all possible combinations of scans.
In this case we found the resolution-based methods to be consistently superior; when the same scan was used both to produce the parcels and to test them (i.e., scans $1$-$1$, $2$-$2$ and $3$-$3$), the ``A'' method is best (lowest unexplained variance), but for the other charts, the Rl and Rr methods had the lowest unexplained variance. 
We further see that the $\ell_2$-method maintains a small advantage over TSVD, which agrees with the superior predictive performance of $\ell_2$-regularization from Fig. \ref{fig_cv_svd_vs_rr}. 
The average for tests of parcels on the same scan of a subject versus their other scans are provided in Table \ref{table_mets_x-y}, where we see consistent performance for the different metrics; resolution methods have the lowest unexplained variance, highest internal correlation, and lowest between-parcel correlations for cross-scan tests.
\begin{table}[]
	\centering
	\caption{Table of metrics for different clustering methods, as measured on same scan that clustering was performed versus measured on different (cross) scan to test generalizability; The resolution-based methods (Rr and Rl) perform consistently better on cross-scan tests with lowest unexplained varience, highest internal correlation, and lowest correlations between parcels.}
	\begin{tabular}{l|cc|cc|cc}
		& \multicolumn{2}{c}{Unexplained Variance} & \multicolumn{2}{c}{Internal Correlation} &  \multicolumn{2}{c}{Parcel Correlation}  \\ 
		Method & Same & Cross & Same & Cross & Same & Cross	\\ \hline
		Rr & 0.340 & 0.358 & 0.653 & 0.635 & 0.468 & 0.493 \\
		Rl & 0.314 & 0.352 & 0.680 & 0.638 & 0.454 & 0.492 \\
		A & 0.291 & 0.370 & 0.704 & 0.620 & 0.447 & 0.504 \\
		Ar & 0.292 & 0.374 & 0.702 & 0.618 & 0.446 & 0.506 \\
		AA & 0.347 & 0.464 & 0.644 & 0.525 & 0.486 & 0.582 \\
		WNN & 0.359 & 0.378 & 0.635 & 0.606 & 0.517 & 0.521 \\
		NN & 0.373 & 0.380 & 0.618 & 0.611 & 0.539 & 0.537 \\
		XYZ & 0.392 & 0.391 & 0.600 & 0.601 & 0.538 & 0.538 \\
		AAL & 0.397 & 0.396 & 0.591 & 0.594 & 0.536 & 0.536 \\
		RNG & 0.654 & 0.653 & 0.333 & 0.334 & 0.961 & 0.961 
		\end{tabular}
	\label{table_mets_x-y}
\end{table}

%
%
%
%
%
%
%

\section{Discussion}

In summary, we demonstrated how the concept of resolution could be adapted to the neighborhood regression problem for estimating network connectivity.
We showed that the intuitive idea of clustering this resolution matrix led to a new kind of spectral clustering. 
Further, we found different algorithm variants depending on the form of regularization used for the neighborhood prediction; 
a SVD truncation-based regularization led to a more traditional algorithm based on clustering of singular vectors, while a $\ell_2$-penalized regularization led to an algorithm based on clustering weighted singular vectors.
This provides a new perspective on spectral clustering which allows more principled decisions on open questions such as the choice of model order as well as on data preprocessing decisions, which are typically made independently and heuristically.

We tested the approach for parcellation of fMRI data, and found that the proposed methods yielded parcels which were more consistent across scans as well as more compact spatially versus conventional approaches based on clustering of voxel time series or their correlations.
Further, while the spatially-constrained spectral clustering method produced parcels with higher dice coefficients and smaller average size, the unexplained variance is higher for this method, meaning the parcels are a worse approximation to the measured data overall.  
It is particularly interesting that the resolution-based methods outperform the basic time-series clustering (the ``A'' method) for the cross-scan tests of unexplained variance, suggesting that their basis for clustering is more robust. 
The fact that the ``A'' method is superior when tested on the same scans provides validation that the method was performed properly. 
Also the fact that the methods WNN, NN, and XYZ, which produce more compact clusters (and indeed slightly more similar clusters), also performed worse on the unexplained variance shows that the success of the resolution-based methods here is not simply due to increased consistency or compactness of clusters.
Further the lack of improvement in the``Ar'' method over the ``A'' method suggests it is not simply a benefit due to the regularization.

As noted above, the homogeneity metrics are biased towards the basic ``A'' method;
it is a method which groups the most similar time series, and the homogeneity metrics test which method successfully grouped the most similar time series. 
A metric which might be more interesting from a hierarchical network perspective is something similar to our cross-validated preprocessing, where we test which parcellation can produce the best network. 
we might measure this by testing how well we can use the signals from other parcels to predict a given parcel's signal.
The difficulty with this metric is that it rewards bad parcellations. If a certain module with high internal connectivity is broken in to multiple parcels, those parcels can be used to more reliably predict their neighbors (which should be in the same parcel). 
This is of course compounded by spatial smoothing which creates correlations between nearby voxels.

%
%
%
%
%

A more efficent method for determining the regularization in the preprocessing stage would also be valuable . 
The exclusion of the local neighborhood (to prevent self-loops from biasing every result towards the unregularized extreme) prevents the exploitation of the low-rank structure of the covariance matrix, as each neighborhood has a different exclusion region.
This renders the approach difficult for very large datasets, and requires subsequent metrics of parcellation or later stages to determine the optimal preprocessing parameters.

\section{acknowledgments} 

The authors wish to thank the NIH (R01 GM109068, R01 MH104680, R01 MH107354) and NSF (1539067) for their partial support.

%
\bibliography{zoterorefs}
\bibliographystyle{plain}

\end{document}